\documentclass[pdflatex,sn-mathphys]{sn-jnl}

\jyear{2023}%

\usepackage{ulem}
\usepackage{verbatim}

\usepackage{mathtools}
\DeclarePairedDelimiter\bra{\langle}{\rvert}
\DeclarePairedDelimiter\ket{\lvert}{\rangle}
\DeclarePairedDelimiterX\braket[2]{\langle}{\rangle}{#1}

\newcommand{\nocomments}{\long\def\comm##1\commend{}}

\nocomments

\newcommand{\akcom}[1]{{\comm \color{red} Akash: ``#1'' \commend}}
\newcommand{\petroscom}[1]{{\comm \color{red} Petros: ``#1'' \commend}}
\newcommand{\johcom}[1]{\comm \textcolor{red}{Johannes: ``{#1}''} \commend}

\newcommand{\ak}[1]{{\color{black} #1}}
\newcommand{\petros}[1]{{\color{black} #1}}
\newcommand{\joh}[1]{{\color{black} #1}}

\newcommand{\aknew}[1]{{\color{black} #1}}
\newcommand{\papnew}[1]{{\color{black} #1}}

\newcommand{\finrev}[1]{{\color{black} #1}}

\begin{document}

\title{Unconventional magnetism mediated by spin-phonon-photon coupling}

%
%

\author*{\fnm{Petros~Andreas~Pantazopoulos, Johannes~Feist, Francisco~J.~Garc\'ia-Vidal, and Akashdeep~Kamra}}\email{petros.pantazopoulos@uam.es; johannes.feist@uam.es; fj.garcia@uam.es; akashdeep.kamra@uam.es}

\affil{\orgdiv{Departamento de F\'isica Te\'orica de la Materia Condensada and Condensed Matter Physics Center (IFIMAC)}, \orgname{Universidad Aut\'onoma de Madrid}, \orgaddress{\city{Madrid}, \postcode{E-28049}, \country{Spain}}}

\abstract{Magnetic order typically emerges due to the short-range exchange interaction between the constituent electronic spins. Recent discoveries have found a crucial role for spin-phonon coupling in various phenomena from optical ultrafast magnetization switching to dynamical control of the magnetic state. 
Here, we demonstrate theoretically the emergence of a biquadratic long-range interaction between spins mediated by their coupling to phonons hybridized with vacuum photons into polaritons. 
The resulting ordered state enabled by the exchange of virtual polaritons between spins is reminiscent of superconductivity mediated by the exchange of virtual phonons. 
The biquadratic nature of the spin-spin interaction promotes ordering without favoring ferro- or antiferromagnetism. 
It further makes the phase transition to magnetic order a first-order transition, unlike in conventional magnets. 
Consequently, a large magnetization develops abruptly on lowering the temperature which \aknew{could} enable magnetic memories admitting ultralow-power thermally-assisted writing while maintaining a high data stability. 
The role of photons in the phenomenon further enables an in-situ static control over the magnetism. 
These unique features make our predicted spin-spin interaction and magnetism highly unconventional paving the way for novel scientific and technological opportunities.}


\maketitle

\section{Introduction}\label{sec:intro}

Magnets host a broad range of intriguing ground states from spin liquids~\cite{Knolle2019} to topological textures~\cite{Yu2021}, such as skyrmions. 
These play a central role in the various correlated states of electronic matter and subfields of physics, like spintronics~\cite{Zutic2004} and unconventional superconductivity~\cite{Stewart2017}. 
Magnets have also had a tremendous impact on contemporary computing technology via magnetic random access memories~\cite{Akerman2005} and read heads based on the magnetoresistance effects~\cite{Fert2008}. 
Different kinds of ordering, such as ferromagnetic and antiferromagnetic, emerge primarily due to the short-range exchange interaction between neighboring spins, which get aligned parallel or antiparallel to each other.

While exchange is typically the strongest interaction in magnets, spin-lattice or spin-phonon coupling has been found to underlie the transfer of spin angular momentum between the magnetic and lattice degrees of freedom~\cite{Bozhko2020}. 
Phenomena such as ultrafast optical switching of magnetic moments~\cite{Stamm2007,Kirilyuk2010,Dornes2019,Maldonado2020,Tauchert2022,Sharma2022} and long-range transport of spin via phonons~\cite{Kittel1949,Weiler2012,Kamra2015,Kikkawa2016,Holanda2018,An2020,Ruckriegel2020} rely fundamentally on the spin-phonon interaction.
\ak{Another of its key consequences is mediating a linear~\cite{Kittel1949,Weiler2012,Kamra2015,Kikkawa2016,Holanda2018,An2020,Ruckriegel2020,Curtis2022} or nonlinear~\cite{Matthews1964,Bittencourt2023,Curtis2022} coupling between phonons and magnons - the spin excitations of ordered magnets. In these considerations, a pre-existing exchange interaction underlies the magnetically ordered ground state while the spin-phonon interaction enables mutual coupling and control between the excitations.
Spin-phonon coupling }has also been exploited to dynamically control the magnetic state via optically driving certain phonon modes~\cite{Forst2011,Forst2015,Afanasiev2021,Sharma2022}, similar to recent schemes employing light as a strong drive to achieve control over magnetic or even nonmagnetic states of matter, such as a superconductor~\cite{Fausti2011}. 

At the same time, modifying existing ordered states or phase transitions by tuning the equilibrium electromagnetic environment is currently a highly desired and pursued goal~\cite{Garcia-Vidal2021,Ashida2020}. \ak{Along these lines, the linear-in-spin coupling with light has been predicted to mediate quadratic spin-spin interactions mimicking antiferromagnetic exchange and stabilizing spin liquid states~\cite{Chiocchetta2021,Bostrom_arxiv}. A paradigmatic work~\cite{Black1977} considered a similar linear-in-pseudospin coupling with phonons to demonstrate a quadratic pseudospin-pseudospin interaction examining its effect on the system's dynamical properties. Such a potential linear-in-spin coupling with phonon displacement is forbidden by time-reversal symmetry~\cite{Kittel1949}.}

Although spin-phonon coupling has recently been established to play an important part in a large number of nonequilibrium spin phenomena, its potential role in determining the fundamental interaction and ground state of a magnet has not been explored. 
Here, we theoretically demonstrate the emergence of an unconventional long-range, \aknew{algebraically-decaying} interaction between localized spins due to the exchange of virtual phonons coupled to the vacuum photon modes. 
The basic phenomenon is similar to how electron-electron attraction emerges in a metal from the exchange of virtual phonons, leading to superconductivity. 
\ak{It is also reminiscent of van der Waals interactions emerging from virtual charge density fluctuations~\cite{chakraborty_next-generation_2020}.}
On account of the spin-phonon coupling \ak{being quadratic-in-spin due to} time-reversal invariance, the emergent spin-spin interaction is found to be biquadratic in the spin components, in contrast with the quadratic nature of the conventional exchange interaction. 
Thus, depending on the sign of the interaction, it enforces order or disorder in the magnet without explicitly favoring parallel or antiparallel configuration.
The possibility to promote disorder can help stabilize spin liquid states~\cite{Knolle2019,Skjaervo2020} at higher temperatures.
Considering the emergence of ferromagnetism i.e., parallel alignment of all spins, the system is found to manifest a first-order phase transition to a large magnetization just below the critical temperature $T_c$. 
Thus, this unconventional magnet enables a promising possibility for memories which would admit ultralow-power thermally-assisted writing~\cite{Thiele2003,Ravelosona2005} of a bit by raising the temperature slightly above $T_c$ and cooling in the presence of a weak applied magnetic field thereby obtaining a large magnetization along a desired direction. 
Such a process becomes ineffective with conventional magnets that manifest a second-order phase transition because cooling slightly below the critical temperature yields a small magnetization. 
Since the latter determines the energy barrier between the two equal-energy bit states, the small magnetization results in an unstable and unreliable data storage.

\section{Emergent spin-spin interaction}\label{sec:interaction}

\begin{figure}[tbh]
\centering
\includegraphics[width=1\linewidth]{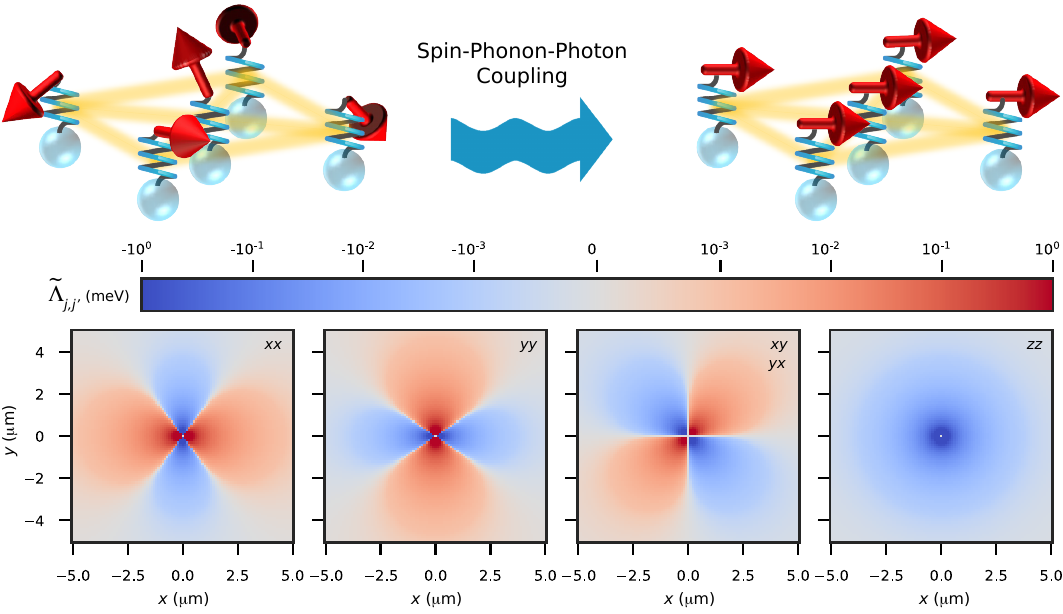}
\caption{{\bf Emergent spin-spin interaction and ordering mediated by spin-phonon-photon coupling.} Top panel: Schematic depiction of the system. 
Localized spins, illustrated as red arrows, are coupled to local phonons, shown as springs. When we consider the phonons to be coupled with global photons (yellow shading) forming polaritonic modes, an exchange of virtual polaritons between the spins causes an effective spin-spin interaction resulting in an in-plane ordering of the spins. 
Bottom panel: Non-vanishing components of the spin-spin coupling tensor $\tilde{\Lambda}_{j,j^\prime}$ between the central nanoparticle located at $\mathbf{r}_j$ and the one situated at $\mathbf{r}_{j^\prime}$ in a square array of spheres in the $x$-$y$ plane.}
\label{fig1}
\end{figure}

We consider ferromagnetic nanoparticles, \ak{each one bearing a large spin with $S \gg 1$ due to its magnetically ordered state and \ak{an infrared (IR)-}active phonon mode with zero wavenumber and THz range frequency} confined to the nanoparticle (see Fig.~\ref{fig1}). 
While the spin-phonon interaction couples these two subsystems, the nanoparticles remain independent of each other at this level. 
Interaction between the different nanoparticles is provided by the electromagnetic photon modes in the system which are delocalized over the entire space and which couple to the \ak{IR-}active phonon in each of the nanoparticles. 
The total Hamiltonian for the system thus becomes
\begin{eqnarray}\label{eq:hamtot}
H & = H_{\mathrm{S}} + H_{\mathrm{P}} + H_{\mathrm{EM}} + H_{\mathrm{S-P}} + H_{\mathrm{EM-P}},
\end{eqnarray}
where $H_{\mathrm{S}}$ describes each of the independent \ak{ferromagnetic nanoparticle spins} and may include contributions from Zeeman coupling or any local magnetic anisotropies. 
$H_{\mathrm{P}}$ captures the phonon along each Cartesian direction on each of the nanoparticles. $H_{\mathrm{EM}}$ accounts for the electromagnetic modes of the environment. 

The spin-phonon coupling is \ak{obtained} as
\petros{\begin{eqnarray}\label{eq:sp}
H_{\mathrm{S-P}} & = \sum_j \sum_{k = x,y,z} b_{k}\dfrac{S_{j;k}^2}{\ak{S^2}}  \left(\beta_{jk}^\dagger + \beta_{jk}\right),
\end{eqnarray}}
\petroscom{up to here we haven't mentioned that the nanoparticles are identical then an index ``j" should appear on the denominator, no?}\akcom{I have mentioned it in the first sentence of the section. Also, it was never explicitly mentioned at what point we start assuming identical nanoparticles. Thus, it is better to assume it from the outset.}
where $j$ runs over all the nanoparticles, $S_{j;k}$ is the $k$th Cartesian component of the nanoparticle $j$'s spin, $\beta_{jk}$ is the annihilation operator of nanoparticle $j$'s phonon mode polarized along the $k$th Cartesian component, and $b_k$ parametrizes the spin-phonon interaction strength. \ak{The form of this coupling has been derived within a simple model in Supplementary Note 1. It is quadratic in spin components due to time-reversal symmetry while the linear coupling to an IR-active phonon necessitates a noncentrosymmetric ferromagnet~\cite{Garst2017,Kanazawa2017}, such as $\mathrm{Cu}_2\mathrm{OSe}\mathrm{O}_3$.}
Equation \eqref{eq:sp} is obtained by quantizing the classical magnetoelastic coupling Hamiltonian~\cite{Kittel1949,Kamra2015}\ak{, appropriately generalized to optical phonons (see Supplementary Note 1),} in terms of the phonon ladder operators. The effective spin-phonon coupling $b_k$ can be expressed in terms of the material parameters (see Supplementary Note 1). 
Much of the unconventional nature of the ensuing spin-spin interaction and magnetism is a direct consequence of the spin-phonon coupling [see Eq.~\eqref{eq:sp}] being invariant under time-reversal, such that it remains the same when $\mathbf{S}$ is replaced by $-\mathbf{S}$. 
As a result, our main results discussed below are independent of the model details. 

The nanoparticle phonon modes couple to the electromagnetic photon modes via the dipole-electric field interaction
\begin{eqnarray}
H_{\mathrm{EM-P}} & = \sum_{n,j} \sum_{k = x,y,z} \mathbf{d}_{jk}\cdot\mathbf{E}_n(\mathbf{r}_j) \alpha_n^\dagger \beta_{jk} + \mathrm{H.c.}, 
\end{eqnarray}
where $n$ runs over all the electromagnetic modes, $\alpha_n$ is the annihilation operator for photon mode $n$ and $\mathbf{E}_n(\mathbf{r})$ is its electric field. 
The position vector of nanoparticle $j$ is $\mathbf{r}_j$, with $d_j$ the dipole moment of its phonon modes, and $\mathbf{d}_{jk}=d_j\widehat{\mathbf{k}}$.
We have further employed the rotating wave~\petros{and long-wavelength} \aknew{(see Supplementary Note 7)} approximations in considering the \ak{coupling between the photons and the zero-wavenumber IR-active phonons.}

Diagonalization of the phonon plus electromagnetic Hamiltonian yields polaritonic modes (see Supplementary Note 2) that are delocalized over the whole system. Consequently, the nanoparticle spins interact with the common polaritons in the environment. We obtain the following effective spin Hamiltonian that describes the exchange of virtual bosons by integrating out the polariton modes employing the path integral framework and evaluating the canonical partition function (see Methods, Supplementary Note 2, and Supplementary Note 3)
\petros{\begin{eqnarray}\label{eq:heff}
H_{\mathrm{eff}} = H_{\mathrm{S}} - \dfrac{1}{S^4}\sum_{j,j^\prime}  \mathbf{S}_j^2 \cdot \tilde{\boldsymbol{\Lambda}}_{j,j^\prime} \cdot\mathbf{S}_{j^\prime}^2,
\end{eqnarray} }
where $\mathbf{S}^2 \equiv \left( S_x^2, S_y^2, S_z ^2 \right)$ denotes a vector made by the spin component squares. 
$\tilde{\boldsymbol{\Lambda}}$ is a $3\times 3$ tensor describing the coupling in units of energy between the different Cartesian components and depends on the composition and coupling of the polaritonic modes to the spins. 
\joh{Treating the full continuum of} electromagnetic modes (see Supplementary Note 4), we obtain
\begin{eqnarray}\label{eq:Lambda}
\tilde{\boldsymbol{\Lambda}}_{j,j^\prime} & = \dfrac{b^2 d_{j} d_{j^\prime}}{2\epsilon_0c^2\hbar^2\Omega^2} \mathrm{Re}\left[\omega^2\tilde{\mathbf{G}}(\mathbf{r}_{j},\mathbf{r}_{j^\prime},\omega)\right]_{\omega=0}\;,
\end{eqnarray}  
where $\Omega$ is the phonon frequency (assumed to be the same for all nanoparticles), $\tilde{\mathbf{G}}$ is the dyadic Green's function of the electromagnetic field, $\epsilon_0$ is the vacuum electric permittivity, $c$ is the vacuum speed of light, and the spin-phonon coupling is assumed to be isotropic, i.e., $b_k = b$ for all $k$.
This elegant formula is not restricted to a particular geometry, and thus it allows for studying complex structures while enabling the design of optimized systems.
The strength of the spin-spin coupling depends on optical, phononic, and magnetoelastic material parameters as well as on the \joh{electrostatic ($\omega=0$) response of the} electromagnetic environment. \joh{Notably, it already acts in free space, and does not rely on the presence of a cavity of any kind, nor on achieving strong coupling between light and matter resonances~\cite{Torma2015}, nor on other resonant effects}.

Equations \eqref{eq:heff} and \eqref{eq:Lambda} constitute one of our main results and demonstrate an emergent interaction between the spins. 
For positive (negative) $\tilde{\Lambda}$ components, energy is minimized by having a large (zero) value of the corresponding spin component squares. 
Thus, for positive values of the components of $\tilde{\Lambda}$, the emergent interaction encourages ordering of the spins without explicitly preferring ferromagnetic or antiferromagnetic configuration. 
This symmetry and degeneracy between the two kinds of ordering may be lifted by additional interactions not explicitly considered here. 
Conversely, a negative value of $\tilde{\Lambda}$ components promotes disorder. 
This may reinforce effects such as geometrical frustration in spin liquids~\cite{Knolle2019} leading to their higher stability. 

Considering a two-dimensional square array of spherical nanoparticles and employing known material parameters (see Supplementary Note 5), we show the various non-vanishing components of $\tilde{\Lambda}$ as a function of the distance between the spins in Fig.~\ref{fig1}. 
A strongly anisotropic nature of the emergent interaction can be seen. 
While the in-plane ($x$-$y$) components of $\tilde{\Lambda}$ are positive on average, thereby supporting an ordered state, the out-of-plane ($z$) component remains negative for all spins. 
Thus, the emergent spin-spin interaction encourages the spins to remain in the plane giving rise to two-dimensional magnetism. 

\section{Mean-field theory of unconventional ferromagnetism}\label{sec:mft}

We now examine the emergence of ferromagnetic order due to the spin-spin interaction derived above. 
To this end, we continue to assume a two-dimensional organization of the nanoparticles in either a square or a hexagonal pattern. 
Further, we consider spherical or cylindrical nanoparticles. 
In the former, phonons polarized along any spatial direction interact equally with the electric field. In contrast, for cylindrical particles, the phonon mode polarized along the cylinder axis will couple most strongly to the electric field, so that the interaction can be approximated as being due to a single dipole component.

\begin{figure}[tbh]
\centering
\includegraphics[width=1\linewidth]{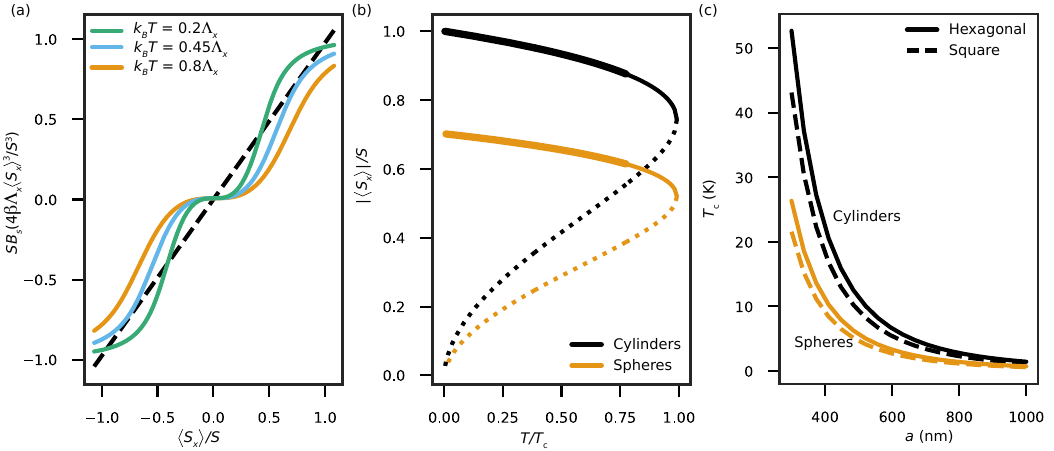}
\caption{{\bf Temperature dependence of the unconventional ferromagnetic state.} (a) Graphical solution of the self-consistency equation for the mean value of spin $x$ component, \petros{$\left \langle S_{x} \right \rangle /S= SB_S\left(4\beta \Lambda_x \langle S_{x} \rangle^3/S^3 \right)$}, for various coupling strengths and \petros{$S\gg1$}.
A solution is obtained when the solid line crosses the dashed line $y=\langle S_{x} \rangle$. \finrev{$k_B$ and $T$ are the Boltzmann constant and temperature, respectively.}
(b) Temperature dependence of the absolute mean value of the $x$ component of the spin for an array of cylindrical (black line) and spherical (orange line) nanoparticles, with solid (dotted) lines indicating locally stable (unstable) solutions. The thick solid lines indicate that the solution furthermore corresponds to the global minimum of the free energy. (c) Dependence of the critical temperature $T_c$ on the lattice constant $a$ of the square (solid line) and hexagonal (dashed line) array of spherical (orange line) or cylindrical (black line) nanoparticles.}
\label{fig2}
\end{figure}

Assuming all spins to be in the same state due to translational invariance and that no external magnetic field is applied, the spin-spin interaction [see Eq.~\eqref{eq:heff}] results in the following mean-field Hamiltonian (see Supplementary Note 6)
\petros{\begin{eqnarray}\label{eq:hmf}
H_\mathrm{MF} &= -\dfrac{1}{S}\sum_{j}\mathbf{h}_{j}\cdot\mathbf{S}_j\;,
\end{eqnarray}
with $\mathbf{h}_{j}= 4(\Lambda_{j;x} \langle S_x \rangle ^3 \widehat{\mathbf{x}} +  \Lambda_{j;y} \langle S_y \rangle ^3 \widehat{\mathbf{y}} + \Lambda_{j;z}\langle S_z \rangle ^3 \widehat{\mathbf{z}})/S^3$ the effective magnetic field.}
Here, $\langle \cdot \rangle$ denotes the expectation value and $\Lambda_{j;x} \equiv \sum_{j^\prime} \tilde{\Lambda}_{j,j^\prime}^{xx}$ and so on. 
Thus, only the diagonal components of the $\tilde{\Lambda}_{ij}$ tensor contribute to the net magnetic order in the whole ensemble. 
We obtain the self-consistency equation for determining the $x$ component of each spin (see Supplementary Note 6) 
\petros{\begin{eqnarray}\label{eq:selfcons}
\left \langle S_{x} \right \rangle & =\dfrac{h_x}{h}SB_S(\beta h)
\end{eqnarray}}
and so on for the $y$ and $z$ components where \petros{$B_S(x)$ is the Brillouin function and} $\beta = 1/ (k_B T)$ with $k_B$ the Boltzmann constant and $T$ the temperature.

Figure~\ref{fig2}(a) qualitatively shows the graphical solution of the self-consistency Eq.~\eqref{eq:selfcons} assuming non-vanishing coupling only along the $x$ axis for \petros{$S\gg1$}, i.e., \petros{$\left \langle S_{x} \right \rangle = SB_s \left(4\beta \Lambda_x \langle S_{x} \rangle^3 /S^3\right)$}.
In contrast with the case of conventional ferromagnetism which admits a unique stable solution to the self-consistency equation, here we find two solutions.
This is a direct result of the mean-field $\mathbf{h}$ components scaling as the third power of the corresponding spin component expectation value, instead of the first power as is the case for conventional magnetism. 

The absolute value of the two solutions as a function of temperature is displayed in Fig~\ref{fig2}(b). 
We find that the spin expectation value develops a finite and large value abruptly as the temperature is lowered, which corresponds to a first-order phase transition. 
In contrast, conventional magnetism corresponds to a second-order phase transition in which the magnetization increases gradually as the temperature is lowered below the Curie temperature. 
The solution associated with high expectation value is stable and the other is unstable, because they correspond to a minimum and a maximum in the Helmholtz free energy, respectively.
When temperature decreases further, the stable solution replaces the trivial solution, $\langle S_x \rangle=0$, as the energetically favorable state since it becomes the global minimum of the free energy.
Furthermore, due to the negative value of $\Lambda_{j;z}$, the expectation value of the spin $z$ component vanishes and the spins are oriented in the $x$-$y$ plane. 

Figure~\ref{fig2}(c) shows the critical temperature of the ferromagnetic state as a function of the lattice constant for the two kinds of nanoparticle arrays considered, in air, thereby providing guidance for achieving a desired critical temperature by choosing the right arrangement of the nanoparticles.
The critical temperature is affected by both the density and the configuration of the lattice. 
As can also be seen in Figs.~\ref{fig2}(b), the shape of the nanoparticles modifies the expectation value of the spin $x$ components and the critical temperature, and can thus be used as an additional degree of freedom to obtain the desired result. 
The reason is the geometrical factor appearing in the self-consistency equation and specifically in $h$.
For the case of cylinders, the effective magnetic field $h$ is proportional to $\Lambda_x$ while for the case of spheres it is proportional to $\sqrt{2}\Lambda_x$ due to equal contributions from $x$ and $y$ components, yielding a lower critical temperature.

\section{Static control of magnetism}\label{sec:control}

There has been a tremendous interest in and technological need for controlling magnetic ground states via external knobs. 
In conventional magnets relying on exchange interaction between the electronic spins, this proves to be a daunting task since it requires controlling the internal states of and interactions between the constituent electrons. 
Nevertheless, transient control over these has been obtained at ultrafast timescales by creating a strong nonequilibrium in the participating electrons~\cite{Golias2021,Forst2011,Forst2015,Afanasiev2021,Sharma2022,Baierl2016,Siegrist2019,Cheng2020,Watzel2022}. 
Since the unconventional magnetism under consideration relies, in part, on the photon modes, it opens the possibility to control the spin interactions and the consequent magnetic order by modifying \ak{these} modes, which are easily accessible to the outside world. 
A recent experiment~\cite{Thomas2021} has already observed an ex-situ change in ferromagnetism via certain resonant effects, distinct from our considerations here. 
We now demonstrate a strong in-situ tunability of the ferromagnet critical temperature via an external knob, such as a gate voltage, that alters the electromagnetic environment of the system. 

\begin{figure}[tbh]
\centering
\includegraphics[width=1\linewidth]{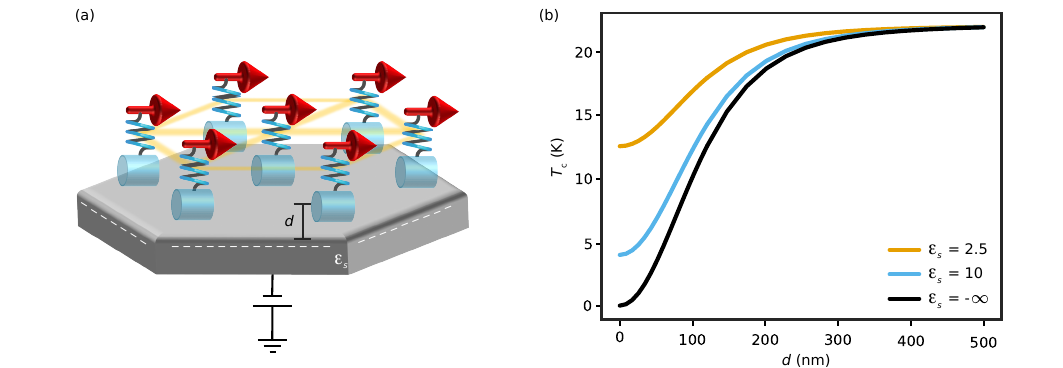}
\caption{ {\bf Control of magnetism via modulation of the substrate dielectric constant $\epsilon_s$.} 
(a) Schematic view of a hexagonal array of cylindrical nanoparticles placed at a distance $d$ from the substrate. 
The semiconducting substrate can  be tuned from being an insulator to a conductor via an applied voltage. The corresponding change in its static dielectric constant alters the polariton modes that mediate the spin-spin interaction. 
(b) Dependence of the critical temperature \finrev{$T_c$} of the unconventional ferromagnetic state on $d$ when the substrate is an electrical insulator ($\epsilon_s = 2.5$; orange line, $\epsilon_s = 10$; blue line) or a conductor (black line).}
\label{fig3}
\end{figure}

To this end, we consider that the nanoparticles are dispersed in an electrically insulating medium with \petros{static permittivity} $\epsilon_m$ (with free space corresponding to $\epsilon_m = 1$). 
Further, the nanoparticles are now deposited on a substrate and placed at distance $d$ from it [see Fig.~\ref{fig3}(a)]. 
The photon modes supported by the system are now modified via the electric field boundary condition imposed by the substrate (see Methods). 
This directly affects the nature of the polaritons mediating the spin-spin interaction and thus the ferromagnetic state. 
Figure \ref{fig3}(b) shows the corresponding critical temperature as a function of nanoparticle-substrate separation for insulating and \petros{\ak{perfectly} conducting} substrates\petros{, where a \ak{critical} temperature decrease is obtained as the nanoparticles approach the substrate.}
\petros{This \ak{is understood by considering} that the spin-spin coupling depends on the Green's function at zero frequency, which can be decomposed into two contributions: the free space and the scattered. 
The latter accounts for the effects due to \ak{the environment engineering}.
\akcom{I do not understand the following statement. Could Johannes please edit and finalize this?} \johcom{How about the following:} \joh{In the case we study here (see Methods), the scattered contribution due to the substrate partially cancels the free-space one, thus decreasing the coupling and consequently the strength of the effective mean field}, which determines the critical temperature (see Eq.~\eqref{eq:selfcons}).}
Interestingly, when the nanoparticles approach the \petros{conducting} substrate, the critical temperature is reduced dramatically and even vanishes in the limit $d\to0$.
\petros{
Intuitively, the underlying mechanism can be understood as follows. The} image dipoles of the particles in the substrate create a field which tends to cancel out the free-space contribution.
Since it is possible to tune, for example, a silicon substrate from being effectively an insulator to a conductor using a gate voltage~\cite{Gong2022}, the critical temperature of the ferromagnetic state may be controlled, in-situ and in equilibrium, via the applied voltage [see Fig.~\ref{fig3}(a)]. 
Furthermore, choosing a specific operating temperature, the gate voltage enables a complete turning on and off of the magnetic order. 
Besides its potential application in memories, this functionality could enable computing architectures based on magnets or spin waves.

\section{Discussion}\label{sec:experimental}
In evaluating the physical observables here, we have employed materials parameters typical of magnetic materials, such as iron garnets and ferrites, as detailed in the Supplementary Note 5.
We note that in order to evaluate the coupling strength of the spin-phonon-photon coupling described here, a candidate material would need to be characterized in terms of both its optical phonons and magnetoelastic properties.
This set of parameters is not available for most materials, presumably in part because this specific combination had not been previously identified as playing an important role.
Therefore, this work implies that an interesting direction for future research would be to look for materials that maximize the effects described here.
For instance, recent experiments demonstrate a significantly enhanced magnon-phonon interaction~\cite{Ramos2019} indicating that our choice of parameters is likely conservative, and better materials would be available.

While we have considered configurations with periodically arranged nanoparticles for concreteness, the qualitative results regarding unconventional ferromagnetism remain the same for arbitrary and disordered two-dimensional arrangements. 
Hence, it is not crucial for experimental realizations to achieve a perfect pattern. 
At the same time, our proposal is valid in and offers a new dimension to artificial spin liquids~\cite{Skjaervo2020} by choosing magnets with optimal optical phonon modes and lithographically fabricating the desired configuration. \ak{Furthermore, our proposed unconventional magnets based on spin-phonon coupling may enable a more effective and faster means of implementing cooling in nanostructures based on adiabatic demagnetization, especially considering the unconventional first-order nature of the transition.}

In conclusion, we theoretically demonstrate an emergent biquadratic spin-spin interaction mediated by spin-phonon-photon coupling.  
The role of photons in this phenomenon lends a strong anisotropy to the spin-spin interaction making the magnetism two-dimensional as well as enabling static in-situ control over the critical temperature of the magnetically ordered state. 
Since our investigated unconventional spin interaction emerges from an exchange of virtual bosons, much like superconductivity, our work opens new avenues for exploring similarly unconventional spin interactions and magnetism mediated by different bosonic modes available in solid state systems.

\section*{Methods}

\subsection*{Obtaining polariton modes}
The $H_\mathrm{P}$, $H_\mathrm{EM}$, and $H_\mathrm{EM-P}$ terms constitute a Hamiltonian which can be represented as
\begin{eqnarray}
H_\mathrm{PP} & = \begin{pmatrix} \boldsymbol{\beta}^\dagger & \boldsymbol{\alpha}^\dagger \end{pmatrix}  \mathbf{H}_\mathrm{PP}
\begin{pmatrix} \boldsymbol{\beta} \\ \boldsymbol{\alpha} \end{pmatrix} \qquad \mathrm{with} \ \ \mathbf{H}_\mathrm{PP}= \begin{pmatrix}
\boldsymbol{\Omega}& \mathbf{g} \\
\mathbf{g}^\dagger& \boldsymbol{\omega} \\
\end{pmatrix},
\end{eqnarray}
where $\boldsymbol{\beta}$ and $\boldsymbol{\alpha}$ are vectors containing the phonon and photon annihilation operators, $\boldsymbol{\Omega}$ and $\boldsymbol{\omega}$ are diagonal matrices describing the energy of the phononic and photonic modes, and $\mathbf{g}$ describes the coupling between the phonons and photons.
The eigenvalues of $\mathbf{H}_\mathrm{PP}$ correspond to the energy of the polaritonic modes, while the eigenvectors contain the coefficients relating the polaritonic operators with the phononic and photonic ones (see Supplementary Note 2).

\subsection*{Integrating out the polaritons}
Equation~\eqref{eq:sp} can be expressed in terms of the polaritons since the phononic operators are related with the polaritonic ones, and thus the total Hamiltonian reduces to $H_\mathrm{S}$, the polaritonic modes and their interaction with spins.
The partition function of the system is calculated in the basis of the spin and polaritonic states.
The effective spin-spin interaction is obtained by calculating the contribution due to  the polaritonic modes employing the path integral framework~\cite{wang_phase_1973,coleman_introduction_2015} (see Supplementary Note 3). 

The coherent state of each polariton is considered. 
Since the polaritons do not interact with each other, their contribution to the partition function is a product of Gaussian integrals (one for each mode), which have analytical solutions.
For each polaritonic mode, the $j$th spin interacts with the $j^\prime$th spin with strength equal to the product of their couplings with the polariton divided by the polariton's energy. 
By summing over all polaritonic modes, the spin-spin coupling can then be related with the inverse of $\mathbf{H}_\mathrm{PP}$. 
For a continuum of modes, the latter can be expressed in terms of the dyadic Green's function of the electromagnetic field (see Supplementary Note 4).

\subsection*{Accounting for a substrate}
The dyadic Green's function of a multilayered structure can be efficiently calculated~\cite{paulus_accurate_2000}.
For a two-layered structure, where the layers are characterized by static electric permittivities $\epsilon_m$ and $\epsilon_s$ and magnetic permeability equal to unity, the coupling reads
\begin{eqnarray*}
	\tilde{\boldsymbol{\Lambda}}_{j,j^\prime}(\mathbf{r}_j,\mathbf{r}_{j^\prime}) & = & \dfrac{d_j d_{j^\prime}}{8\pi\epsilon_0} \dfrac{b^2}{\hbar^2\Omega^2} \left[\tilde{\mathbf{I}}_0(\mathbf{r}_j-\mathbf{r}_{j^\prime}) + \tilde{\mathbf{I}}_{sc}(\lvert{\mathbf{r}_{j;\parallel}}-{\mathbf{r}_{j^\prime;\parallel}}\rvert,\theta,z_j+z_{j^\prime}) \right], \\
	\tilde{\mathbf{I}}_0(\mathbf{r}) & = & \dfrac{1}{ r ^3\epsilon_m}(3\widehat{\mathbf{r}}\otimes\widehat{\mathbf{r}}-\mathbf{1}), \\
	\tilde{\mathbf{I}}_{sc}(R,\theta,z) & = & -\dfrac{\epsilon_m-\epsilon_s}{\epsilon_m+\epsilon_s}\dfrac{2z^2-R^2}{2(z^2+R^2)^{5/2}} 
	\begin{pmatrix} 
			1 & 0 & 0\\	
			0 & 1 & 0\\	
			0 & 0 & 2
	\end{pmatrix} \\
	& &+ \dfrac{\epsilon_m-\epsilon_s}{\epsilon_m+\epsilon_s} \dfrac{3R^2}{2(z^2+R^2)^{5/2}}
	\begin{pmatrix}
		\cos(2\theta)  & \sin(2\theta)  & -2z\cos\theta/R\\
		\sin(2\theta)  & -\cos(2\theta) & -2z\sin\theta/R \\
		2z\cos\theta/R & 2z\sin\theta/R & 0 
	\end{pmatrix},
\end{eqnarray*}
with $\mathbf{r}_j,\mathbf{r}_{j^\prime}$ corresponding to nanoparticles in the $\epsilon_m$ layer.

\backmatter

\bmhead*{Data availability}
All the information required for reproducing these results has been provided in the main text and supplemental information. The numerical data generated for plotting the figures is available on reasonable request. 

\aknew{
	
	\bmhead*{Code availability}
	The numerical routines employed for generating the data plotted in the figures are available on reasonable request. 	
	
}

\bmhead*{Acknowledgments}
We thank Diego Fernandez de la Pradilla \ak{and Mathias Weiler} \akcom{I confirmed with him that our required materials parameters have not been measured for CSO. He does experiments on this material and spin-phonon coupling.} for helpful discussions. 
\finrev{We acknowledge financial support by the Spanish Ministry for Science and Innovation-Agencia Estatal de Investigacion (AEI) through Grants RTI2018-099737-B-I00 (P.A.P.,J.F.,F.J.G.-V.), PID2021-125894NB-I00 (J.F.,F.J.G.-V.), PCI2018-093145374 (through the QuantERA program of the European Commission) (P.A.P.,J.F.,F.J.G.-V.), CEX2018-000805-M (through the Maria de Maeztu program for Units of Excellence in R\&D) (P.A.P.,J.F.,F.J.G.-V.,A.K.) and RYC2021-031063-I (A.K.), by the European Research Council through Grant No. ERC-2016-StG-714870 (P.A.P.,J.F.) and by the European Union’s Horizon Europe Research and Innovation Programme through agreement 101070700 (MIRAQLS, J.F.,F.J.G.-V.). F.J.G.-V. acknowledges financial support from the Comunidad de Madrid and the Spanish State through the Recovery, Transformation and Resilience Plan [``MATERIALES DISRUPTIVOS BIDIMENSIONALES (2D)'' (MAD2D-CM)-UAM7], and the European Union through the Next Generation EU funds.}

\bmhead*{Author contributions}
All authors contributed to the calculations, discussions, and writing of the article.

\bmhead*{Competing interests}
The authors declare that they have no competing interests.

\bibliography{Uncon_mag}


\begin{thebibliography}{77}
\ifx \bisbn   \undefined \def \bisbn  #1{ISBN #1}\fi
\ifx \binits  \undefined \def \binits#1{#1}\fi
\ifx \bauthor  \undefined \def \bauthor#1{#1}\fi
\ifx \batitle  \undefined \def \batitle#1{#1}\fi
\ifx \bjtitle  \undefined \def \bjtitle#1{#1}\fi
\ifx \bvolume  \undefined \def \bvolume#1{\textbf{#1}}\fi
\ifx \byear  \undefined \def \byear#1{#1}\fi
\ifx \bissue  \undefined \def \bissue#1{#1}\fi
\ifx \bfpage  \undefined \def \bfpage#1{#1}\fi
\ifx \blpage  \undefined \def \blpage #1{#1}\fi
\ifx \burl  \undefined \def \burl#1{\textsf{#1}}\fi
\ifx \doiurl  \undefined \def \doiurl#1{\url{https://doi.org/#1}}\fi
\ifx \betal  \undefined \def \betal{\textit{et al.}}\fi
\ifx \binstitute  \undefined \def \binstitute#1{#1}\fi
\ifx \binstitutionaled  \undefined \def \binstitutionaled#1{#1}\fi
\ifx \bctitle  \undefined \def \bctitle#1{#1}\fi
\ifx \beditor  \undefined \def \beditor#1{#1}\fi
\ifx \bpublisher  \undefined \def \bpublisher#1{#1}\fi
\ifx \bbtitle  \undefined \def \bbtitle#1{#1}\fi
\ifx \bedition  \undefined \def \bedition#1{#1}\fi
\ifx \bseriesno  \undefined \def \bseriesno#1{#1}\fi
\ifx \blocation  \undefined \def \blocation#1{#1}\fi
\ifx \bsertitle  \undefined \def \bsertitle#1{#1}\fi
\ifx \bsnm \undefined \def \bsnm#1{#1}\fi
\ifx \bsuffix \undefined \def \bsuffix#1{#1}\fi
\ifx \bparticle \undefined \def \bparticle#1{#1}\fi
\ifx \barticle \undefined \def \barticle#1{#1}\fi
\bibcommenthead
\ifx \bconfdate \undefined \def \bconfdate #1{#1}\fi
\ifx \botherref \undefined \def \botherref #1{#1}\fi
\ifx \url \undefined \def \url#1{\textsf{#1}}\fi
\ifx \bchapter \undefined \def \bchapter#1{#1}\fi
\ifx \bbook \undefined \def \bbook#1{#1}\fi
\ifx \bcomment \undefined \def \bcomment#1{#1}\fi
\ifx \oauthor \undefined \def \oauthor#1{#1}\fi
\ifx \citeauthoryear \undefined \def \citeauthoryear#1{#1}\fi
\ifx \endbibitem  \undefined \def \endbibitem {}\fi
\ifx \bconflocation  \undefined \def \bconflocation#1{#1}\fi
\ifx \arxivurl  \undefined \def \arxivurl#1{\textsf{#1}}\fi
\csname PreBibitemsHook\endcsname

\bibitem{Knolle2019}
\begin{barticle}
\bauthor{\bsnm{Knolle}, \binits{J.}},
\bauthor{\bsnm{Moessner}, \binits{R.}}:
\batitle{A field guide to spin liquids}.
\bjtitle{Annual Review of Condensed Matter Physics}
\bvolume{10}(\bissue{1}),
\bfpage{451}--\blpage{472}
(\byear{2019}).
\doiurl{10.1146/annurev-conmatphys-031218-013401}
\end{barticle}
\endbibitem

\bibitem{Yu2021}
\begin{barticle}
\bauthor{\bsnm{Yu}, \binits{H.}},
\bauthor{\bsnm{Xiao}, \binits{J.}},
\bauthor{\bsnm{Schultheiss}, \binits{H.}}:
\batitle{Magnetic texture based magnonics}.
\bjtitle{Physics Reports}
\bvolume{905},
\bfpage{1}--\blpage{59}
(\byear{2021}).
\doiurl{10.1016/j.physrep.2020.12.004}.
\bcomment{Magnetic texture based magnonics}
\end{barticle}
\endbibitem

\bibitem{Zutic2004}
\begin{barticle}
\bauthor{\bparticle{\ifmmode \check{Z}\else
  \v{Z}\fi{}uti\ifmmode~\acute{c}\else} \bsnm{\'{c}\fi{}}, \binits{I.}},
\bauthor{\bsnm{Fabian}, \binits{J.}},
\bauthor{\bsnm{Das~Sarma}, \binits{S.}}:
\batitle{Spintronics: Fundamentals and applications}.
\bjtitle{Rev. Mod. Phys.}
\bvolume{76},
\bfpage{323}--\blpage{410}
(\byear{2004}).
\doiurl{10.1103/RevModPhys.76.323}
\end{barticle}
\endbibitem

\bibitem{Stewart2017}
\begin{barticle}
\bauthor{\bsnm{Stewart}, \binits{G.R.}}:
\batitle{Unconventional superconductivity}.
\bjtitle{Advances in Physics}
\bvolume{66}(\bissue{2}),
\bfpage{75}--\blpage{196}
(\byear{2017}).
\doiurl{10.1080/00018732.2017.1331615}
\end{barticle}
\endbibitem

\bibitem{Akerman2005}
\begin{barticle}
\bauthor{\bsnm{Åkerman}, \binits{J.}}:
\batitle{Toward a universal memory}.
\bjtitle{Science}
\bvolume{308}(\bissue{5721}),
\bfpage{508}--\blpage{510}
(\byear{2005}).
\doiurl{10.1126/science.1110549}
\end{barticle}
\endbibitem

\bibitem{Fert2008}
\begin{barticle}
\bauthor{\bsnm{Fert}, \binits{A.}}:
\batitle{Nobel lecture: Origin, development, and future of spintronics}.
\bjtitle{Rev. Mod. Phys.}
\bvolume{80},
\bfpage{1517}--\blpage{1530}
(\byear{2008}).
\doiurl{10.1103/RevModPhys.80.1517}
\end{barticle}
\endbibitem

\bibitem{Bozhko2020}
\begin{barticle}
\bauthor{\bsnm{Bozhko}, \binits{D.A.}},
\bauthor{\bsnm{Vasyuchka}, \binits{V.I.}},
\bauthor{\bsnm{Chumak}, \binits{A.V.}},
\bauthor{\bsnm{Serga}, \binits{A.A.}}:
\batitle{Magnon-phonon interactions in magnon spintronics (review article)}.
\bjtitle{Low Temperature Physics}
\bvolume{46}(\bissue{4}),
\bfpage{383}--\blpage{399}
(\byear{2020}).
\doiurl{10.1063/10.0000872}
\end{barticle}
\endbibitem

\bibitem{Stamm2007}
\begin{barticle}
\bauthor{\bsnm{Stamm}, \binits{C.}},
\bauthor{\bsnm{Kachel}, \binits{T.}},
\bauthor{\bsnm{Pontius}, \binits{N.}},
\bauthor{\bsnm{Mitzner}, \binits{R.}},
\bauthor{\bsnm{Quast}, \binits{T.}},
\bauthor{\bsnm{Holldack}, \binits{K.}},
\bauthor{\bsnm{Khan}, \binits{S.}},
\bauthor{\bsnm{Lupulescu}, \binits{C.}},
\bauthor{\bsnm{Aziz}, \binits{E.F.}},
\bauthor{\bsnm{Wietstruk}, \binits{M.}},
\bauthor{\bsnm{D\"urr}, \binits{H.A.}},
\bauthor{\bsnm{Eberhardt}, \binits{W.}}:
\batitle{Femtosecond modification of electron localization and transfer of
  angular momentum in nickel}.
\bjtitle{Nature Materials}
\bvolume{6},
\bfpage{740}
(\byear{2007}).
\doiurl{10.1038/nmat1985}
\end{barticle}
\endbibitem

\bibitem{Kirilyuk2010}
\begin{barticle}
\bauthor{\bsnm{Kirilyuk}, \binits{A.}},
\bauthor{\bsnm{Kimel}, \binits{A.V.}},
\bauthor{\bsnm{Rasing}, \binits{T.}}:
\batitle{Ultrafast optical manipulation of magnetic order}.
\bjtitle{Rev. Mod. Phys.}
\bvolume{82},
\bfpage{2731}--\blpage{2784}
(\byear{2010}).
\doiurl{10.1103/RevModPhys.82.2731}
\end{barticle}
\endbibitem

\bibitem{Dornes2019}
\begin{barticle}
\bauthor{\bsnm{Dornes}, \binits{C.}},
\bauthor{\bsnm{Acremann}, \binits{Y.}},
\bauthor{\bsnm{Savoini}, \binits{M.}},
\bauthor{\bsnm{Kubli}, \binits{M.}},
\bauthor{\bsnm{Neugebauer}, \binits{M.J.}},
\bauthor{\bsnm{Abreu}, \binits{E.}},
\bauthor{\bsnm{Huber}, \binits{L.}},
\bauthor{\bsnm{Lantz}, \binits{G.}},
\bauthor{\bsnm{Vaz}, \binits{C.A.F.}},
\bauthor{\bsnm{Lemke}, \binits{H.}},
\bauthor{\bsnm{Bothschafter}, \binits{E.M.}},
\bauthor{\bsnm{Porer}, \binits{M.}},
\bauthor{\bsnm{Esposito}, \binits{V.}},
\bauthor{\bsnm{Rettig}, \binits{L.}},
\bauthor{\bsnm{Buzzi}, \binits{M.}},
\bauthor{\bsnm{Alberca}, \binits{A.}},
\bauthor{\bsnm{Windsor}, \binits{Y.W.}},
\bauthor{\bsnm{Beaud}, \binits{P.}},
\bauthor{\bsnm{Staub}, \binits{U.}},
\bauthor{\bsnm{Zhu}, \binits{D.}},
\bauthor{\bsnm{Song}, \binits{S.}},
\bauthor{\bsnm{Glownia}, \binits{J.M.}},
\bauthor{\bsnm{Johnson}, \binits{S.L.}}:
\batitle{The ultrafast einstein–de haas effect}.
\bjtitle{Nature}
\bvolume{565},
\bfpage{209}
(\byear{2019}).
\doiurl{10.1038/s41586-018-0822-7}
\end{barticle}
\endbibitem

\bibitem{Maldonado2020}
\begin{barticle}
\bauthor{\bsnm{Maldonado}, \binits{P.}},
\bauthor{\bsnm{Chase}, \binits{T.}},
\bauthor{\bsnm{Reid}, \binits{A.H.}},
\bauthor{\bsnm{Shen}, \binits{X.}},
\bauthor{\bsnm{Li}, \binits{R.K.}},
\bauthor{\bsnm{Carva}, \binits{K.}},
\bauthor{\bsnm{Payer}, \binits{T.}},
\bauthor{\bparticle{Horn~von} \bsnm{Hoegen}, \binits{M.}},
\bauthor{\bsnm{Sokolowski-Tinten}, \binits{K.}},
\bauthor{\bsnm{Wang}, \binits{X.J.}},
\bauthor{\bsnm{Oppeneer}, \binits{P.M.}},
\bauthor{\bsnm{D\"urr}, \binits{H.A.}}:
\batitle{Tracking the ultrafast nonequilibrium energy flow between electronic
  and lattice degrees of freedom in crystalline nickel}.
\bjtitle{Phys. Rev. B}
\bvolume{101},
\bfpage{100302}
(\byear{2020}).
\doiurl{10.1103/PhysRevB.101.100302}
\end{barticle}
\endbibitem

\bibitem{Tauchert2022}
\begin{barticle}
\bauthor{\bsnm{Tauchert}, \binits{S.R.}},
\bauthor{\bsnm{Volkov}, \binits{M.}},
\bauthor{\bsnm{Ehberger}, \binits{D.}},
\bauthor{\bsnm{Kazenwadel}, \binits{D.}},
\bauthor{\bsnm{Evers}, \binits{M.}},
\bauthor{\bsnm{Lange}, \binits{H.}},
\bauthor{\bsnm{Donges}, \binits{A.}},
\bauthor{\bsnm{Book}, \binits{A.}},
\bauthor{\bsnm{Kreuzpaintner}, \binits{W.}},
\bauthor{\bsnm{Nowak}, \binits{U.}},
\bauthor{\bsnm{Baum}, \binits{P.}}:
\batitle{Polarized phonons carry angular momentum in ultrafast
  demagnetization}.
\bjtitle{Nature}
\bvolume{602},
\bfpage{73}
(\byear{2022}).
\doiurl{10.1038/s41586-021-04306-4}
\end{barticle}
\endbibitem

\bibitem{Sharma2022}
\begin{barticle}
\bauthor{\bsnm{Sharma}, \binits{S.}},
\bauthor{\bsnm{Shallcross}, \binits{S.}},
\bauthor{\bsnm{Elliott}, \binits{P.}},
\bauthor{\bsnm{Dewhurst}, \binits{J.K.}}:
\batitle{Making a case for femto-phono-magnetism with fept}.
\bjtitle{Science Advances}
\bvolume{8}(\bissue{37}),
\bfpage{2021}
(\byear{2022}).
\doiurl{10.1126/sciadv.abq2021}
\end{barticle}
\endbibitem

\bibitem{Kittel1949}
\begin{barticle}
\bauthor{\bsnm{Kittel}, \binits{C.}}:
\batitle{Physical theory of ferromagnetic domains}.
\bjtitle{Rev. Mod. Phys.}
\bvolume{21},
\bfpage{541}--\blpage{583}
(\byear{1949}).
\doiurl{10.1103/RevModPhys.21.541}
\end{barticle}
\endbibitem

\bibitem{Weiler2012}
\begin{barticle}
\bauthor{\bsnm{Weiler}, \binits{M.}},
\bauthor{\bsnm{Huebl}, \binits{H.}},
\bauthor{\bsnm{Goerg}, \binits{F.S.}},
\bauthor{\bsnm{Czeschka}, \binits{F.D.}},
\bauthor{\bsnm{Gross}, \binits{R.}},
\bauthor{\bsnm{Goennenwein}, \binits{S.T.B.}}:
\batitle{Spin pumping with coherent elastic waves}.
\bjtitle{Phys. Rev. Lett.}
\bvolume{108},
\bfpage{176601}
(\byear{2012}).
\doiurl{10.1103/PhysRevLett.108.176601}
\end{barticle}
\endbibitem

\bibitem{Kamra2015}
\begin{barticle}
\bauthor{\bsnm{Kamra}, \binits{A.}},
\bauthor{\bsnm{Keshtgar}, \binits{H.}},
\bauthor{\bsnm{Yan}, \binits{P.}},
\bauthor{\bsnm{Bauer}, \binits{G.E.W.}}:
\batitle{Coherent elastic excitation of spin waves}.
\bjtitle{Phys. Rev. B}
\bvolume{91},
\bfpage{104409}
(\byear{2015}).
\doiurl{10.1103/PhysRevB.91.104409}
\end{barticle}
\endbibitem

\bibitem{Kikkawa2016}
\begin{barticle}
\bauthor{\bsnm{Kikkawa}, \binits{T.}},
\bauthor{\bsnm{Shen}, \binits{K.}},
\bauthor{\bsnm{Flebus}, \binits{B.}},
\bauthor{\bsnm{Duine}, \binits{R.A.}},
\bauthor{\bsnm{Uchida}, \binits{K.-i.}},
\bauthor{\bsnm{Qiu}, \binits{Z.}},
\bauthor{\bsnm{Bauer}, \binits{G.E.W.}},
\bauthor{\bsnm{Saitoh}, \binits{E.}}:
\batitle{Magnon polarons in the spin seebeck effect}.
\bjtitle{Phys. Rev. Lett.}
\bvolume{117},
\bfpage{207203}
(\byear{2016}).
\doiurl{10.1103/PhysRevLett.117.207203}
\end{barticle}
\endbibitem

\bibitem{Holanda2018}
\begin{barticle}
\bauthor{\bsnm{Holanda}, \binits{J.}},
\bauthor{\bsnm{Maior}, \binits{D.S.}},
\bauthor{\bsnm{Azevedo}, \binits{A.}},
\bauthor{\bsnm{Rezende}, \binits{S.M.}}:
\batitle{Detecting the phonon spin in magnon–phonon conversion experiments}.
\bjtitle{Nature Physics}
\bvolume{14},
\bfpage{500}
(\byear{2018}).
\doiurl{10.1038/s41567-018-0079-y}
\end{barticle}
\endbibitem

\bibitem{An2020}
\begin{barticle}
\bauthor{\bsnm{An}, \binits{K.}},
\bauthor{\bsnm{Litvinenko}, \binits{A.N.}},
\bauthor{\bsnm{Kohno}, \binits{R.}},
\bauthor{\bsnm{Fuad}, \binits{A.A.}},
\bauthor{\bsnm{Naletov}, \binits{V.V.}},
\bauthor{\bsnm{Vila}, \binits{L.}},
\bauthor{\bsnm{Ebels}, \binits{U.}},
\bauthor{\bparticle{de} \bsnm{Loubens}, \binits{G.}},
\bauthor{\bsnm{Hurdequint}, \binits{H.}},
\bauthor{\bsnm{Beaulieu}, \binits{N.}},
\bauthor{\bsnm{Ben~Youssef}, \binits{J.}},
\bauthor{\bsnm{Vukadinovic}, \binits{N.}},
\bauthor{\bsnm{Bauer}, \binits{G.E.W.}},
\bauthor{\bsnm{Slavin}, \binits{A.N.}},
\bauthor{\bsnm{Tiberkevich}, \binits{V.S.}},
\bauthor{\bsnm{Klein}, \binits{O.}}:
\batitle{Coherent long-range transfer of angular momentum between magnon kittel
  modes by phonons}.
\bjtitle{Phys. Rev. B}
\bvolume{101},
\bfpage{060407}
(\byear{2020}).
\doiurl{10.1103/PhysRevB.101.060407}
\end{barticle}
\endbibitem

\bibitem{Ruckriegel2020}
\begin{barticle}
\bauthor{\bsnm{R\"uckriegel}, \binits{A.}},
\bauthor{\bsnm{Duine}, \binits{R.A.}}:
\batitle{Long-range phonon spin transport in ferromagnet--nonmagnetic insulator
  heterostructures}.
\bjtitle{Phys. Rev. Lett.}
\bvolume{124},
\bfpage{117201}
(\byear{2020}).
\doiurl{10.1103/PhysRevLett.124.117201}
\end{barticle}
\endbibitem

\bibitem{Curtis2022}
\begin{barticle}
\bauthor{\bsnm{Curtis}, \binits{J.B.}},
\bauthor{\bsnm{Grankin}, \binits{A.}},
\bauthor{\bsnm{Poniatowski}, \binits{N.R.}},
\bauthor{\bsnm{Galitski}, \binits{V.M.}},
\bauthor{\bsnm{Narang}, \binits{P.}},
\bauthor{\bsnm{Demler}, \binits{E.}}:
\batitle{Cavity magnon-polaritons in cuprate parent compounds}.
\bjtitle{Phys. Rev. Res.}
\bvolume{4},
\bfpage{013101}
(\byear{2022}).
\doiurl{10.1103/PhysRevResearch.4.013101}
\end{barticle}
\endbibitem

\bibitem{Matthews1964}
\begin{barticle}
\bauthor{\bsnm{Matthews}, \binits{H.}},
\bauthor{\bsnm{Morgenthaler}, \binits{F.R.}}:
\batitle{Phonon-pumped spin-wave instabilities}.
\bjtitle{Phys. Rev. Lett.}
\bvolume{13},
\bfpage{614}--\blpage{616}
(\byear{1964}).
\doiurl{10.1103/PhysRevLett.13.614}
\end{barticle}
\endbibitem

\bibitem{Bittencourt2023}
\begin{barticle}
\bauthor{\bsnm{Bittencourt}, \binits{V.A.S.V.}},
\bauthor{\bsnm{Potts}, \binits{C.A.}},
\bauthor{\bsnm{Huang}, \binits{Y.}},
\bauthor{\bsnm{Davis}, \binits{J.P.}},
\bauthor{\bsnm{Viola~Kusminskiy}, \binits{S.}}:
\batitle{Magnomechanical backaction corrections due to coupling to higher-order
  walker modes and kerr nonlinearities}.
\bjtitle{Phys. Rev. B}
\bvolume{107},
\bfpage{144411}
(\byear{2023}).
\doiurl{10.1103/PhysRevB.107.144411}
\end{barticle}
\endbibitem

\bibitem{Forst2011}
\begin{barticle}
\bauthor{\bsnm{F\"orst}, \binits{M.}},
\bauthor{\bsnm{Tobey}, \binits{R.I.}},
\bauthor{\bsnm{Wall}, \binits{S.}},
\bauthor{\bsnm{Bromberger}, \binits{H.}},
\bauthor{\bsnm{Khanna}, \binits{V.}},
\bauthor{\bsnm{Cavalieri}, \binits{A.L.}},
\bauthor{\bsnm{Chuang}, \binits{Y.-D.}},
\bauthor{\bsnm{Lee}, \binits{W.S.}},
\bauthor{\bsnm{Moore}, \binits{R.}},
\bauthor{\bsnm{Schlotter}, \binits{W.F.}},
\bauthor{\bsnm{Turner}, \binits{J.J.}},
\bauthor{\bsnm{Krupin}, \binits{O.}},
\bauthor{\bsnm{Trigo}, \binits{M.}},
\bauthor{\bsnm{Zheng}, \binits{H.}},
\bauthor{\bsnm{Mitchell}, \binits{J.F.}},
\bauthor{\bsnm{Dhesi}, \binits{S.S.}},
\bauthor{\bsnm{Hill}, \binits{J.P.}},
\bauthor{\bsnm{Cavalleri}, \binits{A.}}:
\batitle{Driving magnetic order in a manganite by ultrafast lattice
  excitation}.
\bjtitle{Phys. Rev. B}
\bvolume{84},
\bfpage{241104}
(\byear{2011}).
\doiurl{10.1103/PhysRevB.84.241104}
\end{barticle}
\endbibitem

\bibitem{Forst2015}
\begin{barticle}
\bauthor{\bsnm{F\"orst}, \binits{M.}},
\bauthor{\bsnm{Caviglia}, \binits{A.D.}},
\bauthor{\bsnm{Scherwitzl}, \binits{R.}},
\bauthor{\bsnm{Mankowsky}, \binits{R.}},
\bauthor{\bsnm{Zubko}, \binits{P.}},
\bauthor{\bsnm{Khanna}, \binits{V.}},
\bauthor{\bsnm{Bromberger}, \binits{H.}},
\bauthor{\bsnm{Wilkins}, \binits{S.B.}},
\bauthor{\bsnm{Chuang}, \binits{Y.-D.}},
\bauthor{\bsnm{Lee}, \binits{W.S.}},
\bauthor{\bsnm{Schlotter}, \binits{W.F.}},
\bauthor{\bsnm{Turner}, \binits{J.J.}},
\bauthor{\bsnm{Dakovski}, \binits{G.L.}},
\bauthor{\bsnm{Minitti}, \binits{M.P.}},
\bauthor{\bsnm{Robinson}, \binits{J.}},
\bauthor{\bsnm{Clark}, \binits{S.R.}},
\bauthor{\bsnm{Jaksch}, \binits{D.}},
\bauthor{\bsnm{Triscone}, \binits{J.-M.}},
\bauthor{\bsnm{Hill}, \binits{J.P.}},
\bauthor{\bsnm{Dhesi}, \binits{S.S.}},
\bauthor{\bsnm{Cavalleri}, \binits{A.}}:
\batitle{Spatially resolved ultrafast magnetic dynamics initiated at a complex
  oxide heterointerface}.
\bjtitle{Nature Materials}
\bvolume{14},
\bfpage{883}
(\byear{2015}).
\doiurl{10.1038/nmat4341}
\end{barticle}
\endbibitem

\bibitem{Afanasiev2021}
\begin{barticle}
\bauthor{\bsnm{Afanasiev}, \binits{D.}},
\bauthor{\bsnm{Hortensius}, \binits{J.R.}},
\bauthor{\bsnm{Ivanov}, \binits{B.A.}},
\bauthor{\bsnm{Sasani}, \binits{A.}},
\bauthor{\bsnm{Bousquet}, \binits{E.}},
\bauthor{\bsnm{Blanter}, \binits{Y.M.}},
\bauthor{\bsnm{Mikhaylovskiy}, \binits{R.V.}},
\bauthor{\bsnm{Kimel}, \binits{A.V.}},
\bauthor{\bsnm{Caviglia}, \binits{A.D.}}:
\batitle{Ultrafast control of magnetic interactions via light-driven phonons}.
\bjtitle{Nature Materials}
\bvolume{20},
\bfpage{607}
(\byear{2021}).
\doiurl{10.1038/s41563-021-00922-7}
\end{barticle}
\endbibitem

\bibitem{Fausti2011}
\begin{barticle}
\bauthor{\bsnm{Fausti}, \binits{D.}},
\bauthor{\bsnm{Tobey}, \binits{R.I.}},
\bauthor{\bsnm{Dean}, \binits{N.}},
\bauthor{\bsnm{Kaiser}, \binits{S.}},
\bauthor{\bsnm{Dienst}, \binits{A.}},
\bauthor{\bsnm{Hoffmann}, \binits{M.C.}},
\bauthor{\bsnm{Pyon}, \binits{S.}},
\bauthor{\bsnm{Takayama}, \binits{T.}},
\bauthor{\bsnm{Takagi}, \binits{H.}},
\bauthor{\bsnm{Cavalleri}, \binits{A.}}:
\batitle{Light-induced superconductivity in a stripe-ordered cuprate}.
\bjtitle{Science}
\bvolume{331}(\bissue{6014}),
\bfpage{189}--\blpage{191}
(\byear{2011}).
\doiurl{10.1126/science.1197294}
\end{barticle}
\endbibitem

\bibitem{Garcia-Vidal2021}
\begin{barticle}
\bauthor{\bsnm{Garcia-Vidal}, \binits{F.J.}},
\bauthor{\bsnm{Ciuti}, \binits{C.}},
\bauthor{\bsnm{Ebbesen}, \binits{T.W.}}:
\batitle{Manipulating matter by strong coupling to vacuum fields}.
\bjtitle{Science}
\bvolume{373}(\bissue{6551}),
\bfpage{0336}
(\byear{2021}).
\doiurl{10.1126/science.abd0336}
\end{barticle}
\endbibitem

\bibitem{Ashida2020}
\begin{barticle}
\bauthor{\bsnm{Ashida}, \binits{Y.}},
\bauthor{\bparticle{\ifmmode \dot{I}\else
  \.{I}\fi{}mamo\ifmmode~\breve{g}\else} \bsnm{\u{g}\fi{}lu}, \binits{A.m.c.}},
\bauthor{\bsnm{Faist}, \binits{J.}},
\bauthor{\bsnm{Jaksch}, \binits{D.}},
\bauthor{\bsnm{Cavalleri}, \binits{A.}},
\bauthor{\bsnm{Demler}, \binits{E.}}:
\batitle{Quantum electrodynamic control of matter: Cavity-enhanced
  ferroelectric phase transition}.
\bjtitle{Phys. Rev. X}
\bvolume{10},
\bfpage{041027}
(\byear{2020}).
\doiurl{10.1103/PhysRevX.10.041027}
\end{barticle}
\endbibitem

\bibitem{Chiocchetta2021}
\begin{barticle}
\bauthor{\bsnm{Chiocchetta}, \binits{A.}},
\bauthor{\bsnm{Kiese}, \binits{D.}},
\bauthor{\bsnm{Zelle}, \binits{C.P.}},
\bauthor{\bsnm{Piazza}, \binits{F.}},
\bauthor{\bsnm{Diehl}, \binits{S.}}:
\batitle{Cavity-induced quantum spin liquids}.
\bjtitle{Nature Communications}
\bvolume{12},
\bfpage{5901}
(\byear{2021}).
\doiurl{10.1038/s41467-021-26076-3}
\end{barticle}
\endbibitem

\bibitem{Bostrom_arxiv}
\begin{botherref}
\oauthor{\bsnm{Boström}, \binits{E.V.}},
\oauthor{\bsnm{Sriram}, \binits{A.}},
\oauthor{\bsnm{Claassen}, \binits{M.}},
\oauthor{\bsnm{Rubio}, \binits{A.}}:
Controlling the magnetic state of the proximate quantum spin liquid
  $\alpha$-RuCl$_3$ with an optical cavity
(2022)
\end{botherref}
\endbibitem

\bibitem{Black1977}
\begin{barticle}
\bauthor{\bsnm{Black}, \binits{J.L.}},
\bauthor{\bsnm{Halperin}, \binits{B.I.}}:
\batitle{Spectral diffusion, phonon echoes, and saturation recovery in glasses
  at low temperatures}.
\bjtitle{Phys. Rev. B}
\bvolume{16},
\bfpage{2879}--\blpage{2895}
(\byear{1977}).
\doiurl{10.1103/PhysRevB.16.2879}
\end{barticle}
\endbibitem

\bibitem{chakraborty_next-generation_2020}
\begin{barticle}
\bauthor{\bsnm{Chakraborty}, \binits{D.}},
\bauthor{\bsnm{Berland}, \binits{K.}},
\bauthor{\bsnm{Thonhauser}, \binits{T.}}:
\batitle{Next-{Generation} {Nonlocal} van der {Waals} {Density} {Functional}}.
\bjtitle{Journal of Chemical Theory and Computation}
\bvolume{16}(\bissue{9}),
\bfpage{5893}--\blpage{5911}
(\byear{2020}).
\doiurl{10.1021/acs.jctc.0c00471}.
\bcomment{Publisher: American Chemical Society}.
Accessed 2023-07-25
\end{barticle}
\endbibitem

\bibitem{Skjaervo2020}
\begin{barticle}
\bauthor{\bsnm{Skj{\ae}rv{\o}}, \binits{S.H.}},
\bauthor{\bsnm{Marrows}, \binits{C.H.}},
\bauthor{\bsnm{Stamps}, \binits{R.L.}},
\bauthor{\bsnm{Heyderman}, \binits{L.J.}}:
\batitle{Advances in artificial spin ice}.
\bjtitle{Nature Reviews Physics}
\bvolume{2},
\bfpage{13}
(\byear{2020}).
\doiurl{10.1038/s42254-019-0118-3}
\end{barticle}
\endbibitem

\bibitem{Thiele2003}
\begin{barticle}
\bauthor{\bsnm{Thiele}, \binits{J.-U.}},
\bauthor{\bsnm{Maat}, \binits{S.}},
\bauthor{\bsnm{Fullerton}, \binits{E.E.}}:
\batitle{Ferh/fept exchange spring films for thermally assisted magnetic
  recording media}.
\bjtitle{Applied Physics Letters}
\bvolume{82}(\bissue{17}),
\bfpage{2859}--\blpage{2861}
(\byear{2003}).
\doiurl{10.1063/1.1571232}
\end{barticle}
\endbibitem

\bibitem{Ravelosona2005}
\begin{barticle}
\bauthor{\bsnm{Ravelosona}, \binits{D.}},
\bauthor{\bsnm{Lacour}, \binits{D.}},
\bauthor{\bsnm{Katine}, \binits{J.A.}},
\bauthor{\bsnm{Terris}, \binits{B.D.}},
\bauthor{\bsnm{Chappert}, \binits{C.}}:
\batitle{Nanometer scale observation of high efficiency thermally assisted
  current-driven domain wall depinning}.
\bjtitle{Phys. Rev. Lett.}
\bvolume{95},
\bfpage{117203}
(\byear{2005}).
\doiurl{10.1103/PhysRevLett.95.117203}
\end{barticle}
\endbibitem

\bibitem{Garst2017}
\begin{barticle}
\bauthor{\bsnm{Garst}, \binits{M.}},
\bauthor{\bsnm{Waizner}, \binits{J.}},
\bauthor{\bsnm{Grundler}, \binits{D.}}:
\batitle{Collective spin excitations of helices and magnetic skyrmions: review
  and perspectives of magnonics in non-centrosymmetric magnets}.
\bjtitle{Journal of Physics D: Applied Physics}
\bvolume{50}(\bissue{29}),
\bfpage{293002}
(\byear{2017}).
\doiurl{10.1088/1361-6463/aa7573}
\end{barticle}
\endbibitem

\bibitem{Kanazawa2017}
\begin{botherref}
\oauthor{\bsnm{Kanazawa}, \binits{N.}},
\oauthor{\bsnm{Seki}, \binits{S.}},
\oauthor{\bsnm{Tokura}, \binits{Y.}}:
Ferromagnetic materials: Noncentrosymmetric magnets hosting magnetic skyrmions
  (adv. mater. 25/2017).
Advanced Materials
\textbf{29}(25)
(2017).
\doiurl{10.1002/adma.201770180}
\end{botherref}
\endbibitem

\bibitem{Torma2015}
\begin{barticle}
\bauthor{\bsnm{Törmä}, \binits{P.}},
\bauthor{\bsnm{Barnes}, \binits{W.L.}}:
\batitle{Strong coupling between surface plasmon polaritons and emitters: a
  review}.
\bjtitle{Reports on Progress in Physics}
\bvolume{78}(\bissue{1}),
\bfpage{013901}
(\byear{2014}).
\doiurl{10.1088/0034-4885/78/1/013901}
\end{barticle}
\endbibitem

\bibitem{Golias2021}
\begin{barticle}
\bauthor{\bsnm{Golias}, \binits{E.}},
\bauthor{\bsnm{Kumberg}, \binits{I.}},
\bauthor{\bsnm{Gelen}, \binits{I.}},
\bauthor{\bsnm{Thakur}, \binits{S.}},
\bauthor{\bsnm{G\"ordes}, \binits{J.}},
\bauthor{\bsnm{Hosseinifar}, \binits{R.}},
\bauthor{\bsnm{Guillet}, \binits{Q.}},
\bauthor{\bsnm{Dewhurst}, \binits{J.K.}},
\bauthor{\bsnm{Sharma}, \binits{S.}},
\bauthor{\bsnm{Sch\"u\ss{}ler-Langeheine}, \binits{C.}},
\bauthor{\bsnm{Pontius}, \binits{N.}},
\bauthor{\bsnm{Kuch}, \binits{W.}}:
\batitle{Ultrafast optically induced ferromagnetic state in an elemental
  antiferromagnet}.
\bjtitle{Phys. Rev. Lett.}
\bvolume{126},
\bfpage{107202}
(\byear{2021}).
\doiurl{10.1103/PhysRevLett.126.107202}
\end{barticle}
\endbibitem

\bibitem{Baierl2016}
\begin{barticle}
\bauthor{\bsnm{Baierl}, \binits{S.}},
\bauthor{\bsnm{Hohenleutner}, \binits{M.}},
\bauthor{\bsnm{Kampfrath}, \binits{T.}},
\bauthor{\bsnm{Zvezdin}, \binits{A.K.}},
\bauthor{\bsnm{Kimel}, \binits{A.V.}},
\bauthor{\bsnm{Huber}, \binits{R.}},
\bauthor{\bsnm{Mikhaylovskiy}, \binits{R.V.}}:
\batitle{Nonlinear spin control by terahertz-driven anisotropy fields}.
\bjtitle{Nature Photonics}
\bvolume{10},
\bfpage{715}
(\byear{2016}).
\doiurl{10.1038/nphoton.2016.181}
\end{barticle}
\endbibitem

\bibitem{Siegrist2019}
\begin{barticle}
\bauthor{\bsnm{Siegrist}, \binits{F.}},
\bauthor{\bsnm{Gessner}, \binits{J.A.}},
\bauthor{\bsnm{Ossiander}, \binits{M.}},
\bauthor{\bsnm{Denker}, \binits{C.}},
\bauthor{\bsnm{Chang}, \binits{Y.-P.}},
\bauthor{\bsnm{Schr\"oder}, \binits{M.C.}},
\bauthor{\bsnm{Guggenmos}, \binits{A.}},
\bauthor{\bsnm{Cui}, \binits{Y.}},
\bauthor{\bsnm{Walowski}, \binits{J.}},
\bauthor{\bsnm{Martens}, \binits{U.}},
\bauthor{\bsnm{Dewhurst}, \binits{J.K.}},
\bauthor{\bsnm{Kleineberg}, \binits{U.}},
\bauthor{\bsnm{M\"unzenberg}, \binits{M.}},
\bauthor{\bsnm{Sharma}, \binits{S.}},
\bauthor{\bsnm{Schultze}, \binits{M.}}:
\batitle{Light-wave dynamic control of magnetism}.
\bjtitle{Nature}
\bvolume{571},
\bfpage{240}
(\byear{2019}).
\doiurl{10.1038/s41586-019-1333-x}
\end{barticle}
\endbibitem

\bibitem{Cheng2020}
\begin{barticle}
\bauthor{\bsnm{Cheng}, \binits{O.H.-C.}},
\bauthor{\bsnm{Son}, \binits{D.H.}},
\bauthor{\bsnm{Sheldon}, \binits{M.}}:
\batitle{Light-induced magnetism in plasmonic gold nanoparticles}.
\bjtitle{Nature Photonics}
\bvolume{14},
\bfpage{365}
(\byear{2020}).
\doiurl{10.1038/s41566-020-0603-3}
\end{barticle}
\endbibitem

\bibitem{Watzel2022}
\begin{barticle}
\bauthor{\bsnm{W\"atzel}, \binits{J.}},
\bauthor{\bparticle{Rebernik Ribi\ifmmode~\check{c}\else} \bsnm{\v{c}\fi{}},
  \binits{P.c.v.}},
\bauthor{\bsnm{Coreno}, \binits{M.}},
\bauthor{\bsnm{Danailov}, \binits{M.B.}},
\bauthor{\bsnm{David}, \binits{C.}},
\bauthor{\bsnm{Demidovich}, \binits{A.}},
\bauthor{\bsnm{Di~Fraia}, \binits{M.}},
\bauthor{\bsnm{Giannessi}, \binits{L.}},
\bauthor{\bsnm{Hansen}, \binits{K.}},
\bauthor{\bparticle{Kru\ifmmode \check{s}\else
  \v{s}\fi{}i\ifmmode~\check{c}\else} \bsnm{\v{c}\fi{}}, \binits{i.c.v.}},
\bauthor{\bsnm{Manfredda}, \binits{M.}},
\bauthor{\bsnm{Meyer}, \binits{M.}},
\bauthor{\bparticle{Miheli\ifmmode~\check{c}\else} \bsnm{\v{c}\fi{}},
  \binits{A.}},
\bauthor{\bsnm{Mirian}, \binits{N.}},
\bauthor{\bsnm{Plekan}, \binits{O.}},
\bauthor{\bsnm{Ressel}, \binits{B.}},
\bauthor{\bsnm{R\"osner}, \binits{B.}},
\bauthor{\bsnm{Simoncig}, \binits{A.}},
\bauthor{\bsnm{Spampinati}, \binits{S.}},
\bauthor{\bsnm{Stupar}, \binits{M.}},
\bauthor{\bparticle{\ifmmode~\check{Z}\else} \bsnm{\v{Z}\fi{}itnik},
  \binits{M.c.v.}},
\bauthor{\bsnm{Zangrando}, \binits{M.}},
\bauthor{\bsnm{Callegari}, \binits{C.}},
\bauthor{\bsnm{Berakdar}, \binits{J.}},
\bauthor{\bsnm{De~Ninno}, \binits{G.}}:
\batitle{Light-induced magnetization at the nanoscale}.
\bjtitle{Phys. Rev. Lett.}
\bvolume{128},
\bfpage{157205}
(\byear{2022}).
\doiurl{10.1103/PhysRevLett.128.157205}
\end{barticle}
\endbibitem

\bibitem{Thomas2021}
\begin{barticle}
\bauthor{\bsnm{Thomas}, \binits{A.}},
\bauthor{\bsnm{Devaux}, \binits{E.}},
\bauthor{\bsnm{Nagarajan}, \binits{K.}},
\bauthor{\bsnm{Rogez}, \binits{G.}},
\bauthor{\bsnm{Seidel}, \binits{M.}},
\bauthor{\bsnm{Richard}, \binits{F.}},
\bauthor{\bsnm{Genet}, \binits{C.}},
\bauthor{\bsnm{Drillon}, \binits{M.}},
\bauthor{\bsnm{Ebbesen}, \binits{T.W.}}:
\batitle{Large enhancement of ferromagnetism under a collective strong coupling
  of ybco nanoparticles}.
\bjtitle{Nano Letters}
\bvolume{21}(\bissue{10}),
\bfpage{4365}--\blpage{4370}
(\byear{2021}).
\doiurl{10.1021/acs.nanolett.1c00973}
\end{barticle}
\endbibitem

\bibitem{Gong2022}
\begin{barticle}
\bauthor{\bsnm{Gong}, \binits{T.}},
\bauthor{\bsnm{Spreng}, \binits{B.}},
\bauthor{\bsnm{Camacho}, \binits{M.}},
\bauthor{\bsnm{Liberal}, \binits{I.n.}},
\bauthor{\bsnm{Engheta}, \binits{N.}},
\bauthor{\bsnm{Munday}, \binits{J.N.}}:
\batitle{Electrically switchable casimir forces using transparent conductive
  oxides}.
\bjtitle{Phys. Rev. A}
\bvolume{106},
\bfpage{062824}
(\byear{2022}).
\doiurl{10.1103/PhysRevA.106.062824}
\end{barticle}
\endbibitem

\bibitem{Ramos2019}
\begin{barticle}
\bauthor{\bsnm{Ramos}, \binits{R.}},
\bauthor{\bsnm{Hioki}, \binits{T.}},
\bauthor{\bsnm{Hashimoto}, \binits{Y.}},
\bauthor{\bsnm{Kikkawa}, \binits{T.}},
\bauthor{\bsnm{Frey}, \binits{P.}},
\bauthor{\bsnm{Kreil}, \binits{A.J.E.}},
\bauthor{\bsnm{Vasyuchka}, \binits{V.I.}},
\bauthor{\bsnm{Serga}, \binits{A.A.}},
\bauthor{\bsnm{Hillebrands}, \binits{B.}},
\bauthor{\bsnm{Saitoh}, \binits{E.}}:
\batitle{Room temperature and low-field resonant enhancement of spin seebeck
  effect in partially compensated magnets}.
\bjtitle{Nature Communications}
\bvolume{10},
\bfpage{5162}
(\byear{2019}).
\doiurl{10.1038/s41467-019-13121-5}
\end{barticle}
\endbibitem

\bibitem{wang_phase_1973}
\begin{barticle}
\bauthor{\bsnm{Wang}, \binits{Y.K.}},
\bauthor{\bsnm{Hioe}, \binits{F.T.}}:
\batitle{Phase {Transition} in the {Dicke} {Model} of {Superradiance}}.
\bjtitle{Physical Review A}
\bvolume{7}(\bissue{3}),
\bfpage{831}--\blpage{836}
(\byear{1973}).
\doiurl{10.1103/PhysRevA.7.831}
\end{barticle}
\endbibitem

\bibitem{coleman_introduction_2015}
\begin{bbook}
\bauthor{\bsnm{Coleman}, \binits{P.}}:
\bbtitle{Introduction to {Many}-{Body} {Physics}}.
\bpublisher{Cambridge University Press},
\blocation{Cambridge}
(\byear{2015}).
\doiurl{10.1017/CBO9781139020916}.
\burl{https://www.cambridge.org/core/books/introduction-to-manybody-physics/B7598FC1FCEE0285F5EC767E835854C8}
\end{bbook}
\endbibitem

\bibitem{paulus_accurate_2000}
\begin{barticle}
\bauthor{\bsnm{Paulus}, \binits{M.}},
\bauthor{\bsnm{Gay-Balmaz}, \binits{P.}},
\bauthor{\bsnm{Martin}, \binits{O.J.F.}}:
\batitle{Accurate and efficient computation of the {Green}'s tensor for
  stratified media}.
\bjtitle{Physical Review E}
\bvolume{62}(\bissue{4}),
\bfpage{5797}--\blpage{5807}
(\byear{2000}).
\doiurl{10.1103/PhysRevE.62.5797}.
\bcomment{Publisher: American Physical Society}.
Accessed 2022-06-01
\end{barticle}
\endbibitem

\bibitem{Chikazumi2009}
\begin{bbook}
\bauthor{\bsnm{Chikazumi}, \binits{S.}}:
\bbtitle{Physics of Ferromagnetism}.
\bsertitle{International Series of Monographs on Physics}.
\bpublisher{OUP Oxford}, \blocation{???}
(\byear{2009}).
\burl{https://books.google.es/books?id=AZVfuxXF2GsC}
\end{bbook}
\endbibitem

\bibitem{Neel1954}
\begin{barticle}
\bauthor{\bsnm{{N\'eel, Louis}}}:
\batitle{Anisotropie magn\'etique superficielle et surstructures
  d'orientation}.
\bjtitle{J. Phys. Radium}
\bvolume{15}(\bissue{4}),
\bfpage{225}--\blpage{239}
(\byear{1954}).
\doiurl{10.1051/jphysrad:01954001504022500}
\end{barticle}
\endbibitem

\bibitem{Mankovsky2022}
\begin{barticle}
\bauthor{\bsnm{Mankovsky}, \binits{S.}},
\bauthor{\bsnm{Polesya}, \binits{S.}},
\bauthor{\bsnm{Lange}, \binits{H.}},
\bauthor{\bsnm{Wei\ss{}enhofer}, \binits{M.}},
\bauthor{\bsnm{Nowak}, \binits{U.}},
\bauthor{\bsnm{Ebert}, \binits{H.}}:
\batitle{Angular momentum transfer via relativistic spin-lattice coupling from
  first principles}.
\bjtitle{Phys. Rev. Lett.}
\bvolume{129},
\bfpage{067202}
(\byear{2022}).
\doiurl{10.1103/PhysRevLett.129.067202}
\end{barticle}
\endbibitem

\bibitem{Thingstad2019}
\begin{barticle}
\bauthor{\bsnm{Thingstad}, \binits{E.}},
\bauthor{\bsnm{Kamra}, \binits{A.}},
\bauthor{\bsnm{Brataas}, \binits{A.}},
\bauthor{\bsnm{Sudb\o{}}, \binits{A.}}:
\batitle{Chiral phonon transport induced by topological magnons}.
\bjtitle{Phys. Rev. Lett.}
\bvolume{122},
\bfpage{107201}
(\byear{2019}).
\doiurl{10.1103/PhysRevLett.122.107201}
\end{barticle}
\endbibitem

\bibitem{Go2019}
\begin{barticle}
\bauthor{\bsnm{Go}, \binits{G.}},
\bauthor{\bsnm{Kim}, \binits{S.K.}},
\bauthor{\bsnm{Lee}, \binits{K.-J.}}:
\batitle{Topological magnon-phonon hybrid excitations in two-dimensional
  ferromagnets with tunable chern numbers}.
\bjtitle{Phys. Rev. Lett.}
\bvolume{123},
\bfpage{237207}
(\byear{2019}).
\doiurl{10.1103/PhysRevLett.123.237207}
\end{barticle}
\endbibitem

\bibitem{Brown1966}
\begin{bbook}
\bauthor{\bsnm{Brown}, \binits{W.F.J.}}:
\bbtitle{Magnetoelastic Interactions}.
\bsertitle{Springer Tracts in Natural Philosophy}.
\bpublisher{Springer}, \blocation{???}
(\byear{1966}).
\burl{https://books.google.es/books?id=-5LAq-Po52wC}
\end{bbook}
\endbibitem

\bibitem{Ashcroft1976}
\begin{bbook}
\bauthor{\bsnm{Ashcroft}, \binits{N.W.}},
\bauthor{\bsnm{Mermin}, \binits{N.D.}}:
\bbtitle{Solid State Physics}.
\bsertitle{HRW international editions}.
\bpublisher{Holt, Rinehart and Winston}, \blocation{???}
(\byear{1976}).
\burl{https://books.google.es/books?id=1C9HAQAAIAAJ}
\end{bbook}
\endbibitem

\bibitem{Anand2005}
\begin{barticle}
\bauthor{\bsnm{Anand}, \binits{L.}},
\bauthor{\bsnm{Gurtin}, \binits{M.E.}},
\bauthor{\bsnm{Lele}, \binits{S.P.}},
\bauthor{\bsnm{Gething}, \binits{C.}}:
\batitle{A one-dimensional theory of strain-gradient plasticity: Formulation,
  analysis, numerical results}.
\bjtitle{Journal of the Mechanics and Physics of Solids}
\bvolume{53}(\bissue{8}),
\bfpage{1789}--\blpage{1826}
(\byear{2005}).
\doiurl{10.1016/j.jmps.2005.03.003}
\end{barticle}
\endbibitem

\bibitem{Nomura2019}
\begin{barticle}
\bauthor{\bsnm{Nomura}, \binits{T.}},
\bauthor{\bsnm{Zhang}, \binits{X.-X.}},
\bauthor{\bsnm{Zherlitsyn}, \binits{S.}},
\bauthor{\bsnm{Wosnitza}, \binits{J.}},
\bauthor{\bsnm{Tokura}, \binits{Y.}},
\bauthor{\bsnm{Nagaosa}, \binits{N.}},
\bauthor{\bsnm{Seki}, \binits{S.}}:
\batitle{Phonon magnetochiral effect}.
\bjtitle{Phys. Rev. Lett.}
\bvolume{122},
\bfpage{145901}
(\byear{2019}).
\doiurl{10.1103/PhysRevLett.122.145901}
\end{barticle}
\endbibitem

\bibitem{Kittel_book}
\begin{bbook}
\bauthor{\bsnm{Kittel}, \binits{C.}},
\bauthor{\bsnm{Fong}, \binits{C.}}:
\bbtitle{Quantum Theory of Solids}.
\bpublisher{John Wiley \& Sons},
\blocation{Toronto}
(\byear{1987})
\end{bbook}
\endbibitem

\bibitem{Feynman_book}
\begin{bbook}
\bauthor{\bsnm{Feynman}, \binits{R.P.}}:
\bbtitle{Statistical Mechanics: A Set Of Lectures}.
\bpublisher{CRC Press},
\blocation{Florida}
(\byear{1998}).
\doiurl{10.1201/9780429493034}
\end{bbook}
\endbibitem

\bibitem{Hepp1973}
\begin{barticle}
\bauthor{\bsnm{Hepp}, \binits{K.}},
\bauthor{\bsnm{Lieb}, \binits{E.H.}}:
\batitle{On the superradiant phase transition for molecules in a quantized
  radiation field: the dicke maser model}.
\bjtitle{Annals of Physics}
\bvolume{76},
\bfpage{360}
(\byear{1973})
\end{barticle}
\endbibitem

\bibitem{Roche2021}
\begin{barticle}
\bauthor{\bsnm{Rom\'an-Roche}, \binits{J.}},
\bauthor{\bsnm{Luis}, \binits{F.}},
\bauthor{\bsnm{Zueco}, \binits{D.}}:
\batitle{Photon condensation and enhanced magnetism in cavity qed}.
\bjtitle{Phys. Rev. Lett.}
\bvolume{127},
\bfpage{167201}
(\byear{2021}).
\doiurl{10.1103/PhysRevLett.127.167201}
\end{barticle}
\endbibitem

\bibitem{roman-roche_effective_2022}
\begin{botherref}
\oauthor{\bsnm{Román-Roche}, \binits{J.}},
\oauthor{\bsnm{Zueco}, \binits{D.}}:
Effective theory for matter in non-perturbative cavity {QED}.
SciPost Physics Lecture Notes,
50
(2022).
\doiurl{10.21468/SciPostPhysLectNotes.50}.
Accessed 2023-05-07
\end{botherref}
\endbibitem

\bibitem{buhmann_dispersion_2012}
\begin{bbook}
\bauthor{\bsnm{Buhmann}, \binits{S.Y.}}:
\bbtitle{Dispersion {Forces} {I}}.
\bsertitle{Springer {Tracts} in {Modern} {Physics}},
vol. \bseriesno{247}.
\bpublisher{Springer},
\blocation{Berlin, Heidelberg}
(\byear{2012}).
\doiurl{10.1007/978-3-642-32484-0}
\end{bbook}
\endbibitem

\bibitem{feist_macroscopic_2022}
\begin{barticle}
\bauthor{\bsnm{Feist}, \binits{J.}},
\bauthor{\bsnm{Fernández-Domínguez}, \binits{A.I.}},
\bauthor{\bsnm{García-Vidal}, \binits{F.J.}}:
\batitle{Macroscopic qed for quantum nanophotonics: emitter-centered modes as a
  minimal basis for multiemitter problems}.
\bjtitle{Nanophotonics}
\bvolume{10}(\bissue{1}),
\bfpage{477}--\blpage{489}
(\byear{2021}).
\doiurl{10.1515/nanoph-2020-0451}
\end{barticle}
\endbibitem

\bibitem{Andrews2018}
\begin{barticle}
\bauthor{\bsnm{Andrews}, \binits{D.L.}},
\bauthor{\bsnm{Jones}, \binits{G.A.}},
\bauthor{\bsnm{Salam}, \binits{A.}},
\bauthor{\bsnm{Woolley}, \binits{R.G.}}:
\batitle{Perspective: {{Quantum Hamiltonians}} for optical interactions}.
\bjtitle{J. Chem. Phys.}
\bvolume{148}(\bissue{4}),
\bfpage{040901}
(\byear{2018}).
\doiurl{10.1063/1.5018399}
\end{barticle}
\endbibitem

\bibitem{novotny_principles_2012}
\begin{bbook}
\bauthor{\bsnm{Novotny}, \binits{L.}},
\bauthor{\bsnm{Hecht}, \binits{B.}}:
\bbtitle{Principles of {Nano}-{Optics}},
\bedition{2}nd edn.
\bpublisher{Cambridge University Press},
\blocation{Cambridge}
(\byear{2012}).
\doiurl{10.1017/CBO9780511794193}
\end{bbook}
\endbibitem

\bibitem{Casimir1948}
\begin{barticle}
\bauthor{\bsnm{Casimir}, \binits{H.B.G.}},
\bauthor{\bsnm{Polder}, \binits{D.}}:
\batitle{The {{Influence}} of {{Retardation}} on the {{London-van}} der {{Waals
  Forces}}}.
\bjtitle{Phys. Rev.}
\bvolume{73}(\bissue{4}),
\bfpage{360}
(\byear{1948}).
\doiurl{10.1103/PhysRev.73.360}
\end{barticle}
\endbibitem

\bibitem{shukla_accurate_2022}
\begin{barticle}
\bauthor{\bsnm{Shukla}, \binits{V.}},
\bauthor{\bsnm{Jiao}, \binits{Y.}},
\bauthor{\bsnm{Lee}, \binits{J.-H.}},
\bauthor{\bsnm{Schr\"oder}, \binits{E.}},
\bauthor{\bsnm{Neaton}, \binits{J.B.}},
\bauthor{\bsnm{Hyldgaard}, \binits{P.}}:
\batitle{Accurate nonempirical range-separated hybrid van der waals density
  functional for complex molecular problems, solids, and surfaces}.
\bjtitle{Phys. Rev. X}
\bvolume{12},
\bfpage{041003}
(\byear{2022}).
\doiurl{10.1103/PhysRevX.12.041003}
\end{barticle}
\endbibitem

\bibitem{comstock_magnetoelastic_1965}
\begin{barticle}
\bauthor{\bsnm{Comstock}, \binits{R.L.}}:
\batitle{Magnetoelastic coupling constants of the ferrites and garnets}.
\bjtitle{Proceedings of the IEEE}
\bvolume{53}(\bissue{10}),
\bfpage{1508}--\blpage{1517}
(\byear{1965}).
\doiurl{10.1109/PROC.1965.4263}.
\bcomment{Conference Name: Proceedings of the IEEE}
\end{barticle}
\endbibitem

\bibitem{coey_magnetism_2001}
\begin{bbook}
\bauthor{\bsnm{Coey}, \binits{J.M.D.}}:
\bbtitle{Magnetism and {Magnetic} {Materials}},
\bedition{1}st edn.
\bpublisher{Cambridge University Press}, \blocation{???}
(\byear{2001}).
\doiurl{10.1017/CBO9780511845000}.
\burl{https://www.cambridge.org/core/product/identifier/9780511845000/type/book}
Accessed 2023-02-17
\end{bbook}
\endbibitem

\bibitem{stancil_spin_2009}
\begin{bbook}
\bauthor{\bsnm{Stancil}, \binits{D.D.}},
\bauthor{\bsnm{Prabhakar}, \binits{A.}}:
\bbtitle{Spin {Waves}: {Theory} and {Applications}}.
\bpublisher{Springer},
\blocation{New York}
(\byear{2009}).
\doiurl{10.1007/978-0-387-77865-5}.
\burl{https://www.springer.com/gp/book/9780387778648}
Accessed 2021-02-11
\end{bbook}
\endbibitem

\bibitem{grunberg_optical_1971}
\begin{barticle}
\bauthor{\bsnm{Grunberg}, \binits{P.}},
\bauthor{\bsnm{Koningstein}, \binits{J.A.}},
\bauthor{\bsnm{Uitert}, \binits{L.G.V.}}:
\batitle{Optical {Phonons} in {Iron} {Garnets}}.
\bjtitle{JOSA}
\bvolume{61}(\bissue{12}),
\bfpage{1613}--\blpage{1617}
(\byear{1971}).
\doiurl{10.1364/JOSA.61.001613}.
\bcomment{Publisher: Optica Publishing Group}.
Accessed 2022-11-30
\end{barticle}
\endbibitem

\bibitem{bernhard_-plane_2002}
\begin{barticle}
\bauthor{\bsnm{Bernhard}, \binits{C.}},
\bauthor{\bsnm{Holden}, \binits{T.}},
\bauthor{\bsnm{Huml\'{i}cek}, \binits{J.}},
\bauthor{\bsnm{Munzar}, \binits{D.}},
\bauthor{\bsnm{Golnik}, \binits{A.}},
\bauthor{\bsnm{Kläser}, \binits{M.}},
\bauthor{\bsnm{Wolf}, \binits{T.}},
\bauthor{\bsnm{Carr}, \binits{L.}},
\bauthor{\bsnm{Homes}, \binits{C.}},
\bauthor{\bsnm{Keimer}, \binits{B.}},
\bauthor{\bsnm{Cardona}, \binits{M.}}:
\batitle{In-plane polarized collective modes in detwinned {YBa2Cu3O6}.95
  observed by spectral ellipsometry}.
\bjtitle{Solid State Communications}
\bvolume{121}(\bissue{2}),
\bfpage{93}--\blpage{97}
(\byear{2002}).
\doiurl{10.1016/S0038-1098(01)00451-3}.
Accessed 2022-01-12
\end{barticle}
\endbibitem

\bibitem{feneberg_ordinary_2018}
\begin{barticle}
\bauthor{\bsnm{Feneberg}, \binits{M.}},
\bauthor{\bsnm{Nixdorf}, \binits{J.}},
\bauthor{\bsnm{Neumann}, \binits{M.D.}},
\bauthor{\bsnm{Esser}, \binits{N.}},
\bauthor{\bsnm{Artús}, \binits{L.}},
\bauthor{\bsnm{Cuscó}, \binits{R.}},
\bauthor{\bsnm{Yamaguchi}, \binits{T.}},
\bauthor{\bsnm{Goldhahn}, \binits{R.}}:
\batitle{Ordinary dielectric function of corundumlike $\alpha-\mathrm{Ga_2O_3}$
  from 40 {meV} to 20 {eV}}.
\bjtitle{Physical Review Materials}
\bvolume{2}(\bissue{4}),
\bfpage{044601}
(\byear{2018}).
\doiurl{10.1103/PhysRevMaterials.2.044601}.
\bcomment{Publisher: American Physical Society}.
Accessed 2022-12-08
\end{barticle}
\endbibitem

\bibitem{Gurevich1996}
\begin{bbook}
\bauthor{\bsnm{Gurevich}, \binits{A.G.}},
\bauthor{\bsnm{Melkov}, \binits{G.A.}}:
\bbtitle{Magnetization Oscillations and Waves}.
\bpublisher{CRC Press},
\blocation{Boca Raton}
(\byear{1996}).
\burl{https://books.google.es/books?id=YgQtSvFIvFQC}
\end{bbook}
\endbibitem

\end{thebibliography}

\newpage

\appendix

\renewcommand{\theequation}{S\arabic{equation}}
\renewcommand{\thefigure}{S\arabic{figure}}
\setcounter{equation}{0}
\setcounter{figure}{0}

\section{Supplementary Note 1: Spin-phonon coupling}\label{sn:spin_photon}

In this note, we derive the spin-phonon coupling term employed in the main text within a simplified and general framework. This also offers guidance with respect to the materials relevant for our proposal. We first motivate the spin-phonon coupling term relevant to our analysis on symmetry grounds. Then, we briefly review the well-established theory of spin-phonon or magnetoelastic coupling for ferromagnets and acoustic phonons~\cite{Kittel1949,Chikazumi2009,Kamra2015}. This analysis is then generalized to the case of lattice with a basis thereby formulating spin-phonon coupling for acoustic and optical phonons in a unified framework. This allows us to establish additional symmetry requirements on the materials that may host our considered spin-phonon coupling term. In addition, the developed framework enables an estimation of the spin-optical phonon coupling based on the more widely available measurements of magnetoelastic constants in magnets~\cite{Chikazumi2009}.

\subsection{General form on time-reversal symmetry grounds}
The term ``spin-phonon coupling'' has been employed to discuss a broad range of distinct effects in the literature. Hence, we must first clarify the particular term that we are interested in and justify its relevance. To this end, we begin by considering the lowest order terms in the combined potential energy density for the spin and phonon system. This approach forms the foundation of conventional magnetoelasticity theory~\cite{Kittel1949}. The lowest order terms are
\petros{\begin{enumerate}
		\item $\sim \widehat{S}_k Q $
		\item $\sim \widehat{S}_k Q^2 $
		\item $\sim \widehat{S}^2_{k} Q $
		\item $\sim \widehat{S}^2_{k} Q^2 $
		\item $\sim \mathbf{\nabla} \widehat{S}_k Q $
		\item $\sim \mathbf{\nabla} \widehat{S}_k Q^2 $
		\item $\sim \left( \mathbf{\nabla} \widehat{S}_k \right)^2 Q $
		\item $\sim \left( \mathbf{\nabla} \widehat{S}_k \right)^2 Q^2 $
\end{enumerate} }
where \petros{$\widehat{S}_k \equiv \widehat{S}_k(\mathbf{r})= S_k(\mathbf{r})/S$} represents the direction cosine of the position dependent spin profile with \petros{$k$} representing a Cartesian coordinate, and $Q$ is the generalized displacement coordinate representing the phonon mode in question. In the above list, terms 1, 2, 5, and 6 are forbidden by time-reversal symmetry requiring the Hamiltonian to remain the same under the substitution \petros{$\widehat{S}_k \to - \widehat{S}_k$}. The terms 5 through 8 can further be considered higher-order since they involve the gradient of the spin direction cosine in a ferromagnetic ground state and are also disregarded, similar to what is done in the conventional magnetoelasticity theory~\cite{Kittel1949}. This leaves terms 3 and 4, with the 4th being higher order than the third. Hence, in this work, we focus on the term 3, \petros{$\sim \widehat{S}^2_k Q $}, the effect of which on the ground state of a spin system has not been considered before. On the other hand, and in contrast with term 3, the term 4 gives rise to a shift in the phonon frequency depending on the magnetic state and has been more widely investigated.

Therefore, the time-reversal symmetry alone warrants that in a strong ferromagnet the leading order effect of spin-phonon coupling is captured by a term \petros{$\sim \widehat{S}^2_k Q$}. However, since we focus on a very specific phonon mode in this work - the zero wavenumber infrared (IR) active optical mode - additional symmetry constraints due to the crystal structure may be present. We examine this in more detail below.

\subsection{Magnetoelasticity theory and spin-pair model}

\begin{figure}[tb]
	\begin{center}
		\includegraphics[width=50mm]{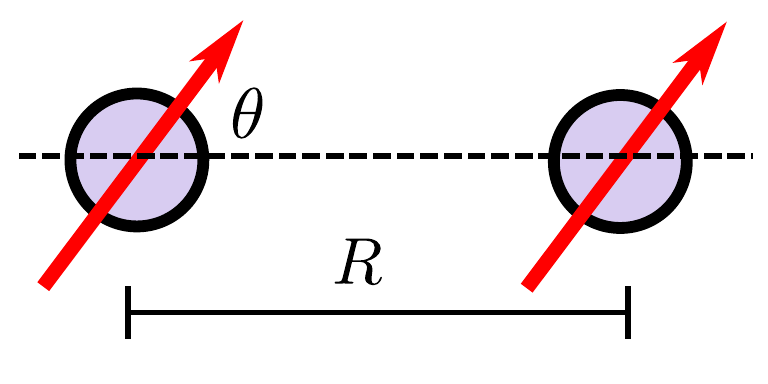}
		\caption{A pair of aligned spins make an angle $\theta$ with the position vector separating them by a distance $R$.}
		\label{fig:spin_pair}
	\end{center}
\end{figure}

Besides relying on time-reversal and crystal symmetry arguments, the conventional magnetoelasticity theory can be constructed using N\'eel's spin-pair model~\cite{Neel1954,Chikazumi2009} depicted in Fig.~\ref{fig:spin_pair}. It considers a pair of aligned spins making an angle $\theta$ with the position vector that separates the spins by a distance $R$. Within a simplified model, it is assumed that the spin-pair interaction energy $W$ depends on $R$ and $\theta$, thereby admitting a Taylor expansion in terms of even spherical harmonics~\cite{Chikazumi2009}:
\begin{align}
	W(R,\theta) & = g(R) + l(R) \left( \cos^2 \theta - \frac{1}{3} \right) + \cdots,
\end{align}
where $g(R)$ and $l(R)$ parametrize the separation dependence of the various terms, and we only show the first two even spherical harmonics. The odd ones are forbidden by, again, invariance under time-reversal. Here, the term $g(R)$ independent of $\theta$ includes contribution from exchange interaction and does not depend on the spin direction. Thus, it does not cause spin-phonon coupling, but instead a renormalization of the elastic forces and equilibrium lattice configuration. We are not interested in this effect here and thus may approximate
\begin{align}\label{eq:wapprox}
	W(R,\theta) & \approx l(R) \cos^2 \theta.
\end{align}
The physical origin of this spin-lattice coupling term is material-dependent with contributions from spin-orbit interaction causing single-ion anisotropies, magnetic dipolar fields and so on~\cite{Kittel1949,Chikazumi2009,Mankovsky2022}. 

Considering how a strain affects the distance $R$ and direction $\theta$ associated with the spin-pair, one may evaluate the resulting change in $W$. Summing over all the spin-pairs in the unit cell, one obtains an expression of the form:
\petros{\begin{align}\label{eq:mec1}
		{H}_\mathrm{mec} & =\int\mathrm{d}\mathbf{r}  \sum_{k = x,y,z} b_1  \dfrac{{S}_k^2}{S^2} u_{kk} +\int\mathrm{d}\mathbf{r}   \sum_{k,k^\prime=x,y,z} b_2 \dfrac{{S}_k {S}_{k^\prime}}{S^2} u_{kk^\prime},
\end{align}}
for the magnetoelastic energy in a cubic crystal~\cite{Kittel1949,Chikazumi2009}. Here, \petros{$u_{kk^\prime} \equiv 1/2 ( \partial R_k / \partial x_{k^\prime} + \partial R_{k^\prime} /\partial x_k)$} are strain tensor components, $b_{1,2}$ parametrize the spin-phonon coupling, \aknew{and ${S}_k \equiv {S}_k(\mathbf{r})$ are the components of spatially resolved spin. In ferromagnetic nanoparticles, the so-called macrospin approximation is valid due to the internal exchange being strong. Thus, the spin components do not depend on the position within the nanoparticle and the integral in the equation above simply yields a volume factor thereby enabling an adequate description in terms of the total nanoparticle spin or its direction (see Sec.~\ref{sec:estimate}). The parameters $b_{1,2}$} can be evaluated in terms of the function $l(R)$ as shown below, but are typically treated as experimentally determined material parameters.

In considering acoustic phonons, one simply replaces the strain components in the equation \eqref{eq:mec1} above by the relevant phonon displacements~\cite{Kamra2015}. While the above has been evaluated for cubic crystals and acoustic phonons, the main form and features of the relations go well-beyond. They also work for optical phonons~\cite{Thingstad2019,Go2019} as well as polycrystalline materials~\cite{Weiler2012}. This is because the main physics relates to how the spin-pair energies are affected by rotation and extension/compression, which can be treated on a more generic footing~\cite{Brown1966}. Due to similar generality reasons, the same expression Eq.~\eqref{eq:mec1} has also been employed, successfully and extensively, for multisublattice ferrimagnets~\cite{Kikkawa2016}, such as yttrium iron garnet, despite their highly complex unit cell.

\subsection{Diatomic chain model}

In this section, we develop a description of spin-phonon coupling within a simple diatomic chain model~\cite{Ashcroft1976} that explicitly accounts for two sublattices and optical phonon modes. A key goal is to use the spin-pair model in addressing the spin-phonon coupling for acoustic and optical phonon branches within a unified framework. Our simplified model is meant to establish and guide general symmetry and phenomenological considerations for multisublattice magnets, which are not easily treated analytically. 

\begin{figure}[tb]
	\begin{center}
		\includegraphics[width=100mm]{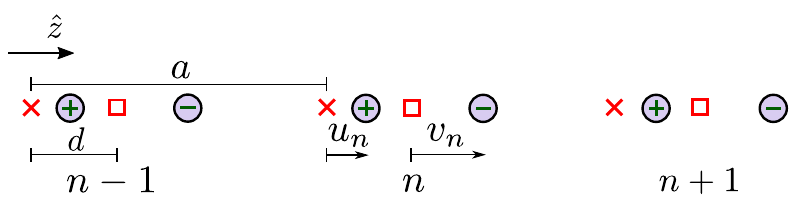}
		\caption{A diatomic chain model for a two-sublattice magnet hosting acoustic and polar optical phonons~\cite{Ashcroft1976}. The equilibrium locations of the two kinds of atoms (circles) are marked by red crosses and squares. Here, $a$ is the lattice constant and $d$ is distance between the two atoms in the basis. The atomic displacements from their respective equilibrium positions are described by $u_n$ and $v_n$.}
		\label{fig:diatomic}
	\end{center}
\end{figure}

The model~\cite{Ashcroft1976}, depicted in Fig.~\ref{fig:diatomic}, consists of a one-dimensional lattice with a basis comprising two oppositely charged atoms displaced by a distance $d$. As a result, the effective spring constants between the neighboring atoms are different depending on the bond distance being $d$ or $a - d$ resulting in the potential energy:
\begin{align}
	U & = \sum_{n} \frac{1}{2} K \left( u_n - v_n \right)^2 +  \frac{1}{2} G \left( u_{n+1} - v_n \right)^2,
\end{align}
where $K$ and $G$ are the two effective spring constants. Assuming the same mass $M$ for the two atoms for simplicity, we obtain the dynamical equations of motion:
\begin{align}
	M \ddot{u}_{n} & = -\frac{\partial U}{\partial u_n} = - K (u_n - v_n) - G (u_n - v_{n-1}), \label{eq:eom1} \\
	M \ddot{v}_{n} & = -\frac{\partial U}{\partial v_n} = - K (v_n - u_n) - G (v_n - u_{n+1}). \label{eq:eom2}
\end{align}
Considering periodic boundary conditions, we assume plane wave solutions of the form
\petros{\begin{align}
		u_{n} = \Re \left( \tilde{u} e^{i \left(q n a - \omega t \right)}  \right),
\end{align}}
and similar for $v_n$. Substituting these in the equations of motion \eqref{eq:eom1} and \eqref{eq:eom2}, we obtain the eigenspectrum and eigenmodes:
\petros{\begin{align}
		\omega^2(q) & = \frac{K + G}{M} \pm \frac{\sqrt{(K - G)^2 + 4 K G \cos^2(q a / 2) }}{M}, \label{eq:spec} \\
		\frac{\tilde{u}}{\tilde{v}} & = \mp \frac{K + G e^{- i q a}}{\sqrt{(K - G)^2 + 4 K G \cos^2(q a / 2)}}. \label{eq:eigmodes}
\end{align}}
The lower (upper) sign in the above expressions corresponds to the acoustic (optical) branch that yields in-phase $\tilde{u} = \tilde{v}$ (out-of-phase $\tilde{u} = - \tilde{v}$) displacements of the atoms for \petros{$q = 0$} mode. Furthermore, it can be seen that substituting $K = G$ (corresponding to $d = a/2$) effectively results in a single phonon branch that is continuously connected at the Brillouin zone boundary \petros{$ q = \pm \pi/a$}. This is expected from a single-sublattice system the eigenmodes of which have been written down using our two-sublattice notation and thus a smaller Brillouin zone.

When considering a spatially homogeneous electric field $E \hat{z}$, i.e., within the long-wavelength approximation as applicable for nanoparticles smaller than the relevant wavelengths, the equations of motion \eqref{eq:eom1} and \eqref{eq:eom2} are modified as follows
\petros{\begin{align}
		M \ddot{u}_{n} &  = - K (u_n - v_n) - G (u_n - v_{n-1}) + Q_a E,  \\
		M \ddot{v}_{n} & = - K (v_n - u_n) - G (v_n - u_{n+1}) - Q_a E,
\end{align}}
where \petros{$Q_a$} is the charge magnitude for both atoms. This results in the electric field coupling to $u_{n} - v_{n}$ and consequently the \petros{$q = 0$} optical mode that corresponds to $\tilde{u} = - \tilde{v}$ [Eq.~\eqref{eq:eigmodes}]. Hence, for coupling to IR light within our polar phonon model, we are primarily interested in the \petros{$q = 0$} mode of the optical branch. 

Having established the phonon eigenmodes, we turn to evaluating the spin-pair energies assuming, as before, that all the atomic spins are aligned. The spin energy [Eq.~\eqref{eq:wapprox}] involving the two spin-pairs is
\begin{align}
	W_S & = \sum_n l_K(R_K) \cos^2 \theta_{K} + l_G(R_G) \cos^2 \theta_{G},
\end{align} 
where the subscripts $K$ and $G$ label the two bonds within a unit cell. The spin-pairs' distances will be affected by longitudinal phonon modes along the z direction while transverse phonons influence $\theta_{K,G}$. Here, we focus on the longitudinal phonon modes and thus, treat $\theta_K = \theta_G = \theta$ as static variables resulting in $\cos \theta$ becoming the third directional cosine of the spin: \petros{$\widehat{S}_z$}. The spin energy may thus be expressed as:
\petros{\begin{align}
		W_S = \sum_n \widehat{S}_z^2 \left( l_K(d) + l_G(a-d) \right) + \widehat{S}_z^2 \left[ l_K^{\prime} (v_n - u_n) + l_G^\prime (u_{n+1} - v_n) \right] ,
\end{align} }
where we have employed $R_K = d + v_n - u_n$ and $R_G = a + u_{n+1} - d - v_n = a - d + u_{n+1} - v_n$ (see Fig.~\ref{fig:diatomic}), Taylor expanded $l_{K,G}$ functions retaining only the first order terms in the phonon displacements, and employed the concise notation $ l_K^{\prime}(d) \equiv l_K^{\prime}$ and $l_G^{\prime}(a - d) \equiv l_G^{\prime}$. Disregarding the constant offset stemming from the equilibrium configuration, we obtain the spin-phonon coupling energy $\Delta W_S$ as:
\petros{\begin{align}\label{eq:deltaw}
		\Delta W_S & = \sum_n \widehat{S}_{z}^2 \left[ l_K^{\prime} (v_n - u_n) + l_G^\prime (u_{n+1} - v_n) \right],
\end{align}}
where we need to express the atomic displacements in terms of the phonons.

Let us first consider the acoustic branch and examine phonons with small, but finite, wavenumber \petros{$q$} such that \petros{$q a \ll 1$}. Employing the lower sign in Eq.~\eqref{eq:eigmodes} and assuming \petros{$u_n = u_0 \cos (q a n)$}, we obtain
\petros{\begin{align}
		\frac{\tilde{v}}{\tilde{u}} & = 1 + \frac{i G q a}{K + G}, \\
		v_n - u_n & = - \frac{G}{K + G} q a u_0 \sin(qan), \\
		u_{n+1} - v_n & =  - \frac{K}{K + G} q a u_0 \sin(qan).
\end{align}}
Employing these relations in Eq.~\eqref{eq:deltaw}, we obtain the spin-phonon energy
\petros{\begin{align}\label{eq:wsacous}
		\Delta W_S & = \sum_n - \left( l_K^\prime \frac{G}{K + G} + l_G^\prime \frac{K}{K + G} \right) q a u_0 \sin (q a n) \widehat{S}_z^2,
\end{align}}
which is our desired result. Since \petros{$ q a u_0 \sin (q a n) \sim \partial u(z) / \partial z$} in a continuum model, Eq.~\eqref{eq:wsacous} validates Eq.~\eqref{eq:mec1} for longitudinal phonons with wavevector along z direction. It further allows us to obtain a simplified microscopic relation $b_1 \propto  l_K^\prime G/(K + G) + l_G^\prime K/(K + G) $. We can also see now why the simple single-sublattice model that would assume $K = G$ and $l_K^\prime = l_G^\prime \equiv l^\prime$ works well for two(multi)-sublattice systems as it will simply imply $b_1 \propto l^\prime$ without changing the overall coupling form. Assuming $K \sim G$, the estimated value of $b_1$ is also obtained correctly. 

Equipped with this understanding of acoustic phonons, let us examine the optical phonon branch. Since we are interested in the IR active optical mode at \petros{$q = 0$}, we disregard even the first order terms in \petros{$q$} here. Considering the upper sign in Eq.~\eqref{eq:eigmodes} and assuming \petros{$u_n = u_0 \cos (q a n)$} as before
\petros{\begin{align}
		\frac{\tilde{v}}{\tilde{u}} & = - 1 , \\
		v_n - u_n & = - 2 u_0 \cos (q n a) = - 2 u_0, \\
		u_{n+1} - v_n & = 2 u_0 \cos (q n a) = 2 u_0 ,
\end{align}}
which, employing Eq.~\eqref{eq:deltaw}, yields 
\petros{\begin{align}\label{eq:wsopt}
		\Delta W_S & = \sum_n  \left( l_G^\prime - l_K^\prime  \right) 2 u_0 \widehat{S}_z^2,
\end{align}}
for the spin-phonon coupling. Assuming $  \lvert l_G^\prime \rvert \ll \lvert l_K^\prime \rvert $, we notice that the spin-phonon coupling above can be obtained from its acoustic phonon counterpart Eq.~\eqref{eq:wsacous} by substituting \petros{$q \sim 2/a$}. This relation and procedure is advantageous/useful since the parameter $b_1$ that enters Eqs.~\eqref{eq:mec1} and \eqref{eq:wsacous} is easily obtained from experiments in a large variety of magnets. On the other hand, to the best of our knowledge, the corresponding parameter for the spin-optical phonon coupling has not been investigated in detail.

\subsection{Consideration of inversion symmetry}

In the previous subsection, we have related the strength of spin-optical phonon coupling to its acoustic counterpart based on the assumption $  \lvert l_G^\prime \rvert \ll \lvert l_K^\prime \rvert $ and within the diatomic model introduced in Ref.~\cite{Ashcroft1976} for studying optical phonons. In reality, we need a weaker condition $  \lvert l_G^\prime - l_K^\prime \rvert \sim l_K^\prime $ for this estimate. Such a condition is ensured by the different bond distances $d$ and $a -d$  in our considered model. We notice that this necessarily breaks the inversion symmetry. While in our model the distances $d$ and $a-d$ need to be different for the existence of optical phonon modes, this is not generically true since we could have obtained the second phonon branch by simple assuming different masses for the two atoms. On the other hand, to justify $l_G^\prime \neq l_K^\prime$ between the two same spins, we necessarily need the two spin bonds to be different thereby breaking inversion symmetry. Thus, we conclude that our considered coupling between the spin and \petros{$q = 0$} IR active phonon is only available in noncentrosymmetric magnets. 

Building on these microscopic insights, we now consider a direct argument to arrive at the same conclusion. Assuming spatial inversion symmetry (due to the crystal structure), only a phonon mode that is odd under spatial inversion may couple to the spatially homogeneous electric field associated with IR light. This is because such a field is odd under spatial inversion while we need the light-phonon coupling term to be invariant under spatial inversion to respect the crystal symmetry. Such a phonon mode cannot directly couple to our assumed aligned spin since the latter is invariant under spatial inversion. Any such coupling between the odd-under-inversion phonon and even-under-inversion spin will violate the inversion symmetry of the crystal.

\subsection{Materials}

Thus, our considered spin-IR active phonon coupling term is available in noncentrosymmetric magnets. We can thus consider materials with a crystal structure that violates inversion symmetry. At the same time, centrosymmetric materials in the bulk can still become noncentrosymmetric under a strain-gradient~\cite{Anand2005} that may be deliberate or accidental. For example, the surface of a ferromagnetic nanoparticle is expected to be highly and nonuniformly strained. Along these lines, buckling of thin layers has also been proposed to convert a centrosymmetric material into a noncentrosymmetric one thereby activating additional couplings~\cite{Curtis2022}.  

Returning to noncentrosymmetric magnets, a large number of candidates exist and some of them have been studied intensely in recent years due to their hosting skyrmions~\cite{Garst2017,Kanazawa2017}. For our considered nanoparticles, the magnetic ground state of each will remain uniform and without any skyrmions, in consistence with our assumptions. One prominent material example is $\mathrm{Cu}_2\mathrm{OSe}\mathrm{O}_3$, in which a phonon magnetochiral effect resulting from Dzyaloshinskii-Moriya interaction has already been observed~\cite{Nomura2019}. However, the particular spin-phonon coupling parameter relevant to our proposal has not been probed, to the best of our knowledge. 

Finding all the optical and spin-phonon parameters relevant to our proposal for a single material has proven to be a daunting task since this specific term of spin-phonon coupling has not been studied much. Also, there is little overlap between such magnetoelastic and IR studies. Hence, in our considerations below, we choose typical parameter values for different materials that have been measured in experiments. We hope that our theoretical proposal will motivate first-principles calculations and/or experiments to characterize the relevant noncentrosymmetric magnets with a focus on this proposal.

\subsection{Estimating the spin-phonon coupling strength}\label{sec:estimate}
As discussed and motivated above, Eq.~\eqref{eq:mec1} for acoustic phonons can be employed for their \petros{$q = 0$} optical counterpart since the latter can be visualized as \petros{$q = \pi/a$} acoustic phonons in an extended Brillouin zone scheme, where $a$ is the lattice constant. Below, we consider the optically-active longitudinal phonon mode polarized along the $z$ direction.
The differently polarized phonons can be considered in an analogous manner. Further, we disregard any transverse phonon modes here.

As per Ref.~\cite{Kittel_book}, the effective longitudinal strain can be related with the canonical position $w(\mathbf{r})$ by $u_{zz}(\mathbf{r})=2w(\mathbf{r})/(a\rho)$, where $\rho$ is the material density.
Quantizing the system as detailed in Ref.~\cite{Feynman_book}, we directly obtain
\petros{\begin{equation}
	u_{zz}(\mathbf{r}) = \dfrac{1}{a}\int \dfrac{\mathrm{d}\mathbf{q}}{\left( 2 \pi\right)^3} \sqrt{\dfrac{2\hbar}{ \rho\Omega(\mathbf{q})}} \left[ {\beta}(\mathbf{q}) e^{i \mathbf{q} \cdot \mathbf{r}} + \ak{{\beta}^{\dagger}(\mathbf{q})} e^{- i \mathbf{q} \cdot \mathbf{r}} \right]\;,
\end{equation}}
where \petros{$\Omega(\mathbf{q})$} is the phonon frequency. 
The operators \petros{${\beta}(\mathbf{q})$ and ${\beta}^\dagger(\mathbf{q})$} pertain to a continuum description. 
Within a small nanoparticle, only a uniform mode is optically active. We thus express the quantities in terms of discrete \petros{$\mathbf{q}$} values and keep only the \petros{$\mathbf{q} = 0$ }mode, obtaining

{\begin{align}\label{eq:uzzfin}
	{u}_{zz} & = \dfrac{1}{a} \sqrt{\dfrac{2\hbar }{\rho V \Omega}} \left( {\beta} + {\beta}^\dagger  \right),
\end{align}}
where $V$ is the volume of the nanoparticle.
Therefore, the interaction between the spin and the phonon operators with the help of Eqs.~\eqref{eq:mec1} and~\eqref{eq:uzzfin}, reads
\petros{\begin{align}
	{H}_{S-P}&= b\dfrac{{S}_z^2}{S^2}  \left( {\beta} + {\beta}^{\dagger}  \right) \;,
\end{align} }
where $b=b_1\sqrt{{2\hbar V }/({\rho \Omega} a^2)} $ quantifies the spin-phonon interaction and has units of energy.

\section{Supplementary Note 2: Polaritonic coupling}\label{sn:pol_cplg} 

Equation (1) of the main text can be written in the form ${H} = H_\mathrm{S} + H_\mathrm{S-P}+ H_\mathrm{PP}$ with
\begin{align}\label{eq:H_PP}
	H_\mathrm{PP} = H_\mathrm{P} + H_\mathrm{EM} + H_\mathrm{EM-P} = \begin{pmatrix} \boldsymbol{\beta}^\dagger & \boldsymbol{\alpha}^\dagger \end{pmatrix}
	\mathbf{H}_\mathrm{PP}
	\begin{pmatrix} \boldsymbol{\beta} \\ \boldsymbol{\alpha} \end{pmatrix}.
\end{align}
with $\mathbf{H}_\mathrm{PP}$ as defined in Eq.~(8) of the main text.
Considering $\mathbf{U}=\begin{pmatrix} \mathbf{C} \\ \mathbf{A}\end{pmatrix}$ to be the unitary matrix that diagonalizes $\mathbf{H}_\mathrm{PP}$, $\mathbf{U}^\dagger \mathbf{H}_\mathrm{PP} \mathbf{U} = \boldsymbol{\tilde{\omega}}$, and defining $\boldsymbol{\pi} = \mathbf{U}^\dagger \begin{pmatrix} \boldsymbol{\beta} \\ \boldsymbol{\alpha} \end{pmatrix}$, $H_\mathrm{PP}$ can be rewritten as
\begin{equation}
	H_\mathrm{PP} = \sum_m\hbar\tilde{\omega}_m\pi_m^\dagger\pi_m \label{eq:H_pol}\;,
\end{equation}
with $\pi_m$, $\pi_m^\dagger$ the creation and annihilation operators of the $m$-th polaritonic mode.
Then, the total Hamiltonian can be written in the form
\petros{\begin{equation}
		{H} =H_\mathrm{S} +\sum_m\hbar\tilde{\omega}_m\pi_m^\dagger\pi_m+ \sum_{j,m,k}b_k\dfrac{S_{j;k}^2}{\ak{S^2}} \big[(C_{jm}^{k})^\star\pi_{m}+C_{jm}^{k}\pi^\dagger_{m}\big] \label{eq:H_tot_pol}\;,
\end{equation}}
where $C_{jm}^{k}$ relate the polaritonic operators to the phononic ones.

With the help of Supplementary Note~3, the polaritonic operators can be integrated out and an effective spin-spin interaction is obtained from Eq.~\eqref{eq:H_tot_pol}, quantified by the coupling strength 
\begin{equation}
	\tilde{\boldsymbol{\Lambda}}_{j,j^\prime}= \Re\big(b^2\mathbf{C}_j{\boldsymbol{\tilde{\omega}}}^{-1}\mathbf{C}_{j^\prime}^\dagger\big)\;.\label{eq:lambda_def}
\end{equation} 

\petros{Now, we can express the spin-spin coupling through the inverse of $\mathbf{H}_\mathrm{PP}$. 
	Using the explicit form of the matrix $\mathbf{H}_\mathrm{PP}$
	\begin{equation}
		\mathbf{H}_\mathrm{PP}= \begin{pmatrix} \boldsymbol{\Omega} & \mathbf{g} \\
			\mathbf{g}^\dagger& \boldsymbol{\omega} \end{pmatrix}
	\end{equation}
	its inverse can be written as
	\begin{equation}\label{eq:Hinv}
		\mathbf{H}_\mathrm{PP}^{-1} = \begin{pmatrix}
			(\boldsymbol{\Omega}-\mathbf{g}\boldsymbol{\omega}^{-1}\mathbf{g}^\dagger)^{-1} & -(\boldsymbol{\Omega}-\mathbf{g}\boldsymbol{\omega}^{-1}\mathbf{g}^\dagger)^{-1} \mathbf{g}\boldsymbol{\omega}^{-1} \\
			-(\boldsymbol{\omega}-\mathbf{g}^\dagger\boldsymbol{\Omega}^{-1}\mathbf{g})^{-1}\mathbf{g}^\dagger\boldsymbol{\Omega}^{-1}& (\boldsymbol{\omega}-\mathbf{g}^\dagger\boldsymbol{\Omega}^{-1}\mathbf{g})^{-1} \end{pmatrix}\;.
	\end{equation}
	Since $\mathbf{U} \boldsymbol{\tilde{\omega}}^{-1} \mathbf{U}^\dagger = \mathbf{H}_\mathrm{PP}^{-1}$ and $\mathbf{U}=(\mathbf{C}\; \mathbf{A})^\mathrm{T}$, we obtain with the help of Eq.~\eqref{eq:Hinv}
	\begin{align}
		\tilde{\boldsymbol{\Lambda}}_{j,j^\prime}&= \Re\Big\{b^2\big[(\boldsymbol{\Omega}-\boldsymbol{\xi})^{-1}\big]_{jj^\prime}\Big\}\label{eq:lambda_long}\;,
	\end{align}
}
where 
\begin{equation}
	\boldsymbol{\xi}=\mathbf{g}\boldsymbol{\omega}^{-1}\mathbf{g}^\dagger = \begin{pmatrix}
		\boldsymbol{\xi}_{11}& \boldsymbol{\xi}_{12} & \ldots &\boldsymbol{\xi}_{1N_p} \\
		\boldsymbol{\xi}_{21}& \boldsymbol{\xi}_{22} & \ldots &\boldsymbol{\xi}_{2N_p} \\
		\vdots  & \vdots& \vdots &       \vdots   \\
		\boldsymbol{\xi}_{N_p1}& \boldsymbol{\xi}_{N_p2} & \ldots &\boldsymbol{\xi}_{N_pN_p} \\
	\end{pmatrix}\label{gog_matrix}\;,
\end{equation}
with $N_p$ indicating the total number of nanoparticles and
\begin{align}
	\boldsymbol{\xi}_{jj^\prime}=d_jd_{j^\prime}\sum_{n=1}^{N_\mathrm{cav}}\dfrac{\mathbf{E}_n(\mathbf{r}_j)\otimes\mathbf{E}_n^\star(\mathbf{r}_{j^\prime})}{\hbar\omega_n}.\label{eq:xi_coupling}
\end{align}
$\otimes$ denotes the dyadic product between two vectors and $N_\mathrm{cav}$ indicates the total number of electromagnetic modes.

\petros{We note that $\boldsymbol{\Omega}$ has dimensions $3N_p \times 3N_p$, $\mathbf{g}$ has dimensions $3N_p \times M$, $\boldsymbol{\omega}$ has dimensions $M\times M$, and  $\tilde{\boldsymbol{\Lambda}}_{j,j^\prime}$ has dimensions $3\times3$, where $M = N_\mathrm{cav} + 3N_p$.}

\section{Supplementary Note 3: Effective spin-spin Hamiltonian}\label{sn:Heff}

An effective spin-spin coupling is derived, based on the spin-phonon-photon interaction, by tracing out the polaritonic degrees of freedom. 
To do so, we employ the theoretical framework detailed in \petros{Refs.~\cite{Hepp1973,wang_phase_1973,Roche2021,coleman_introduction_2015,roman-roche_effective_2022}, which is based on the Euclidean path integral formulation.}  

The thermodynamic functions can be calculated from the canonical partition function $Z=\mathrm{Tr}[\exp(-\beta H)]$ with $\beta = 1/(k_B T)$, $H$ the total Hamiltonian of the system and $k_B$ the Boltzmann constant.
A convenient basis to calculate the trace of the partition function of the bosonic polaritonic modes is formed by the coherent states $\ket{\alpha}$, which are eigenstates of the annihilation operator $a$, i.e., $a\ket{\alpha}=\alpha\ket{\alpha}$, and form a complete set, $\frac{1}{\pi} \int \mathrm{d}^2 \alpha\ket{\alpha}\bra{\alpha}=1$, where the integral is over real and imaginary parts, $\int \mathrm{d}^2 \alpha = \iint \mathrm{d}\!\Re[\alpha]\,\mathrm{d}\!\Im[\alpha]$.

Considering the total Hamiltonian to be of the form
\begin{equation}
	H= H_\mathrm{S} + \sum_{i}\hbar\omega_i a_i^\dagger a_i +\sum_{i,j} \lambda S_j^2(\gamma_{ji}a_i+\gamma_{ji}^\star a_i^\dagger)\;.
\end{equation}
and assuming that there are $N$ spins and $M$ polaritonic modes we have 
\begin{multline}
	Z = \frac{1}{\pi^{M}} \sum_{s_1}\ldots\sum_{s_N}\int \mathrm{d}^2 \alpha_1\ldots\int \mathrm{d}^2 \alpha_M \\ 
	\times \bra{s_1\ldots s_N}\bra{\alpha_1\ldots\alpha_M}\exp(-\beta H)\ket{\alpha_1\ldots\alpha_M}\ket{s_1\ldots s_N}\;,
\end{multline}
where the sum is taken over all spin states. 
\ak{By tracing out the polaritonic degrees of freedom, we wish to obtain an effective Hamiltonian such that 
	\begin{equation}
		Z= \sum_{s_1}\ldots\sum_{s_N}\bra{s_1\ldots s_N}\exp(-\beta H_\mathrm{eff})\ket{s_1\ldots s_N}\;,
	\end{equation}
	with 
	\begin{equation}
		\exp(-\beta H_\mathrm{eff}) = Z_\alpha=\frac{1}{\pi^{M}}\int \mathrm{d}^2 \alpha_1\ldots\int \mathrm{d}^2 \alpha_M
		\bra{\alpha_1\ldots\alpha_M}\exp(-\beta H)\ket{\alpha_1\ldots\alpha_M}\;.\label{eq:Heff_def_many}
	\end{equation}
	In order to accomplish this tracing out \joh{of} the bosonic modes, we employ the result~\cite{wang_phase_1973,coleman_introduction_2015,Roche2021,roman-roche_effective_2022}
	\begin{equation}
		\!\!\!\!\bra{\alpha_i}\exp(-\beta H)\ket{\alpha_i}\!\approx\!\exp\Big\{-\beta\Big[H_\mathrm{S} +\hbar\omega_i \alpha_i^\star \alpha_i +\sum_{j}\lambda S_j^2(\gamma_{ji}\alpha_i+\gamma_{ji}^\star \alpha_i^\star)\Big]\Big\}\;.\label{eq:meanH_alpha_full_many}
	\end{equation}
	and pause to discuss it. 
	Equation \eqref{eq:meanH_alpha_full_many} above is tantamount to replacing the operators $a_i$ by their expectation values $\alpha_i$ in the coherent state. 
	In this sense, it effectively disregards the quantum commutations between operators and is reminiscent of an analogous replacement procedure within the path integral framework for integrating out excitations~\cite{coleman_introduction_2015}. 
	Strictly speaking, Eq.~\eqref{eq:meanH_alpha_full_many} yields the exact result for the free energy per spin in the thermodynamic limit, i.e., when there are a large number of spins ($N \to\infty$). 
	It has been justified semi-rigorously for a finite number $M$ of the bosonic modes~\cite{Hepp1973,wang_phase_1973}. 
	However, since the number of spins is never infinity, one can consider and employ Eq.~\eqref{eq:meanH_alpha_full_many} as an approximation that introduces a usually small error depending on $N$. 
	In this spirit, Eq.~\eqref{eq:meanH_alpha_full_many} has been successfully and widely employed for capturing the essential physics with a good enough accuracy~\cite{wang_phase_1973,Hepp1973,Roche2021,roman-roche_effective_2022}. 
	We also follow this procedure here.
}
Since the modes are not interacting $ Z_\alpha = \exp{(-\beta H_\mathrm{S})}\prod_{i}Z_{\alpha_i}$
with 
\begin{equation}
	Z_{\alpha_i}= \frac{1}{\pi}\int \mathrm{d}^2 \alpha_i \exp\Big\{-\beta\Big[\hbar\omega_i \alpha_i^\star \alpha_i +\sum_{j}\lambda S_j^2(\gamma_{ji}\alpha_i+\gamma_{ji}^\star \alpha_i^\star)\Big]\Big\}\;.
\end{equation}
The integrals have Gaussian form and we get
\begin{align}
	Z_{\alpha_i} = \dfrac{1}{{\beta\hbar\omega_i}}\exp\Big[\beta\sum_{j, j'}\lambda S_j^2\dfrac{\Re(\gamma_{ji}\gamma_{j'i}^\star)}{\hbar\omega_i} \lambda S_{j'}^2\Big]\;.
\end{align}
Then, the total partition function $Z_\alpha$ takes the form
\begin{align}
	Z_{\alpha} =& \exp(-\beta H_\mathrm{S})\prod_{i}\Big\{\dfrac{1}{{\beta\hbar\omega_i}}\exp\Big[\beta\sum_{j, j'}\lambda S_j^2\dfrac{\Re(\gamma_{ji}\gamma_{j'i}^\star)}{\hbar\omega_i} \lambda S_{j'}^2\Big]\Big\}\;.
\end{align}
In the thermodynamic limit~\cite{wang_phase_1973,Roche2021}, we obtain the effective Hamiltonian
\begin{align}
	H_\mathrm{eff} &=H_\mathrm{S}-\sum_{j,j'}{S}_j^2\tilde{\Lambda}_{j,j'}{S}_{j'}^2 \nonumber\\
	\tilde{\Lambda}_{j,j'} &= \Re[\lambda^2\boldsymbol{\gamma}_j\boldsymbol{\omega}^{-1}\boldsymbol{\gamma}_{j'}^\dagger] \label{eq:Heff_full_many}\;,
\end{align}
where $\boldsymbol{\omega}=\mathrm{diag}(\hbar\omega_1,\hbar\omega_2,\ldots,\hbar\omega_M)$, and $\boldsymbol{\gamma}$ is a vector of length $M$ with elements $\gamma_{ji}$. 
We note that the generalization for a spin operator with three components is straightforward, with the coupling $\tilde{\Lambda}$ becoming a $3\times3$ tensor. 

\petroscom{Shall we add a comment of for the Baker-Cambell-Hausdorff formula as a reply to the Reviewer?} \akcom{I do not see the need, but I leave it on Johannes to decide.}\johcom{I do not see the need either.}

\section{Supplementary Note 4: Polaritonic coupling for continuum of EM modes}\label{sn:mqed}
In the derivation \joh{above}, a discrete number of modes has been considered. 
Here, we generalize the analysis for a continuum of electromagnetic modes. 

We follow a macroscopic quantum electrodynamic treatment~\cite{buhmann_dispersion_2012,feist_macroscopic_2022}, in which the quantized electric field is expressed as 
\begin{equation}
	{\mathbf{E}} (\mathbf{r})= \sum_{\lambda}\int_0^\infty \mathrm{d}\omega\int \mathrm{d}^3r'\tilde{\mathbf{G}}_\lambda(\mathbf{r}, \mathbf{r}', \omega){\mathbf{f}}_\lambda( \mathbf{r}', \omega) + \mathrm{H.c.}\label{eq:efield_qed}\;,
\end{equation}
with ${\mathbf{f}}_\lambda(\mathbf{r}', \omega)$ the bosonic annihilation operators of the EM modes, $\lambda$ an index labeling the electric and magnetic contributions and $\tilde{\mathbf{G}}_\lambda(\mathbf{r}, \mathbf{r}', \omega)$ functions related to the (classical) dyadic Green's function of Maxwell's equations.
The electromagnetic interaction of the EM fields with the nanoparticle dipoles \joh{in Coulomb gauge within the Power-Zienau-Woolley picture~\cite{Andrews2018} and using the long-wavelength approximation} is
\begin{equation}
	{H}_\mathrm{EM-P}=\sum_{j}\sum_{\lambda}\int_0^\infty \mathrm{d}\omega\int \mathrm{d}^3r'\mathbf{d}_{j}\cdot[\tilde{\mathbf{G}}_\lambda(\mathbf{r}, \mathbf{r}', \omega){\mathbf{f}}_\lambda( \mathbf{r}', \omega)]\beta_j+\mathrm{H.c.}\;.\label{eq:Hemp_qed}
\end{equation}
By formally discretizing the continuum of cavity modes and expressing it through a collective index $n=\{\lambda,l,\mathbf{r'},\omega\}$ (where $l$ denotes Cartesian components), Eq.~\eqref{eq:xi_coupling} can be expressed in terms of the Green's functions 
\begin{align}
	\xi_{j,j^\prime}^{kk'} &= \sum_{n} \dfrac{g_{jn}^{k}g_{nj^\prime}^{k'\star}}{\hbar\omega_n}= d_{i}d_{j} \sum_i \dfrac{1}{\hbar\omega_i}\sum_{\lambda, l}\int\mathrm{d}^3r' G_\lambda^{kl}(\mathbf{r}_j,\mathbf{r}',\omega_i)[G_\lambda^{k'l}(\mathbf{r}_{j^\prime},\mathbf{r}',\omega_i)]^{\star\mathrm{T}}\nonumber\\
	&=d_{j}d_{j^\prime}\sum_{i}\dfrac{\omega_i }{\pi\epsilon_0c^2}\Im G^{kk'}(\mathbf{r}_j,\mathbf{r}_{j^\prime},\omega_i),
\end{align}
where we have used the property $\sum_{\lambda}\int \mathrm{d}^3r' \tilde{\mathbf{G}}_\lambda(\mathbf{r}_j,\mathbf{r}',\omega)(\tilde{\mathbf{G}}_\lambda(\mathbf{r}_{j^\prime},\mathbf{r}',\omega))^{\star\mathrm{T}}=\hbar\omega^2\Im\tilde{\mathbf{G}}(\mathbf{r}_j,\mathbf{r}_{j^\prime},\omega)/(\pi\epsilon_0c^2)$, where $\epsilon_0$ is the vacuum electric permittivity and $c$ the speed of light in vacuum. \petros{We note that we assumed that each particle has three equivalent dipole (phonon) directions with the same dipole moment $d_j$}. After taking the continuum limit, we obtain
\begin{equation}
	\papnew{\tilde{\boldsymbol{\xi}}}_{j,j^\prime} = \dfrac{d_{j}d_{j^\prime}}{\pi\epsilon_0c^2}\int_0^\infty \mathrm{d}\omega \omega \Im\tilde{\mathbf{G}}(\mathbf{r}_j,\mathbf{r}_{j^\prime},\omega)
	\label{eq:xi_gf}\;.
\end{equation}

To analytically evaluate the integral, we use $\Im z = \frac{z - z^\star}{2i}$ together with the property $\tilde{\mathbf{G}}^\star(\mathbf{r}_i,\mathbf{r}_j,\omega) = \tilde{\mathbf{G}}(\mathbf{r}_i,\mathbf{r}_j,-\omega)$, giving
\begin{equation}
	\int_0^\infty \mathrm{d}\omega\omega \Im\tilde{\mathbf{G}}(\mathbf{r}_i,\mathbf{r}_j,\omega)= \dfrac{1}{2i} \int_{-\infty}^\infty \mathrm{d}\omega\omega \tilde{\mathbf{G}}(\mathbf{r}_i,\mathbf{r}_j,\omega)\;.
\end{equation}
The function $\omega \tilde{\mathbf{G}}(\mathbf{r}_i,\mathbf{r}_j,\omega)$ has a
simple pole at $\omega=0$, and no other poles on the real axis or upper complex half space~\cite{novotny_principles_2012,buhmann_dispersion_2012}. Contour
integration then yields the residue at $\omega=0$, i.e., $\int_{-\infty}^\infty \mathrm{d}\omega\omega \tilde{\mathbf{G}}(\mathbf{r}_i,\mathbf{r}_j,\omega) = i\pi \left[ \omega^2
\tilde{\mathbf{G}}(\mathbf{r}_i,\mathbf{r}_j,\omega)\right]_{\omega=0}$, such
that the polaritonic coupling of Eq.~\eqref{eq:xi_gf} reads
\begin{equation}
	\papnew{\tilde{ \boldsymbol{\xi}}}_{j,j^\prime} = \dfrac{d_jd_{j^\prime}}{2\epsilon_0c^2} \left[\omega^2\tilde{\mathbf{G}}(\mathbf{r}_j,\mathbf{r}_{j^\prime},\omega)\right]_{\omega=0}\label{eq:xi_final}\;.
\end{equation}
\joh{We note that this expression diverges for $j=j'$ due to the singular nature of the free-space Green's function, i.e., due to the divergence of the dipole self-energy. However, this divergence is due to the fact that the long-wavelength approximation means that the nanoparticles behave as point dipoles, and a more careful evaluation would lead to a (small) finite value~\cite{Casimir1948,buhmann_dispersion_2012}. Since these terms just induce a constant energy shift, we instead assume that their contribution is already included in the bare system parameters and discard them in the following. Consequently,}
for couplings smaller than the phonon energy and for  $j\neq j^\prime$, from Eq.~\eqref{eq:lambda_long} we obtain
\begin{align}
	\tilde{\boldsymbol{\Lambda}}_{j,j^\prime} &=\dfrac{b^2 d_j d_{j^\prime}}{2\epsilon_0c^2\hbar^2\Omega^2} \Re\left[\omega^2\tilde{\mathbf{G}}(\mathbf{r}_j,\mathbf{r}_{j^\prime},\omega)\right]_{\omega=0}\;.\label{eq:coupling_ij_final}
\end{align} 

\petros{It is worth noting that before obtaining the analytical expression of the coupling, Eq.~\eqref{eq:coupling_ij_final}, it has been expressed as an integral over all frequencies (see Eq.~\eqref{eq:xi_gf}).
	This characteristic \ak{offers} a link with the van den Waals materials. These are described by nonlocal correlation energy functionals, which are expressed in terms of a kernel and the electron density. 
	The former can be expressed as a frequency integral over all plasmon frequencies and is based on the virtual charge-density fluctuations of the electron gas~\cite{chakraborty_next-generation_2020}. \akcom{Something is missing here.}
	\petroscom{I prepared this based on the two references. Perhaps one of the seniors should check it since none of us is familiar with the field. }
	It can be interpreted that the electron and the associated exchange-correlation hole form an antenna of charged parts~\cite{shukla_accurate_2022}. 
	Similarly, in our case, it can be interpreted that the phonons act as antennas.}

\akcom{A small excerpt of the above discussion needs to appear in the main text. I can do it myself when the missing thing above has been placed.}

\section{Supplementary Note 5: Parameters}\label{sn:params}
For our calculations we consider typical values of magnetic materials. 
\petros{Specifically, the magnetoelastic constant of EuIG $b_1=10^{6}~\mathrm{J/m^3}$~\cite{comstock_magnetoelastic_1965}, density $\rho=5.4~\mathrm{g/cm^3}$~\cite{coey_magnetism_2001,stancil_spin_2009} of YIG, \ak{typical lattice constant} $a=1.1~\mathrm{nm}$~\cite{coey_magnetism_2001,stancil_spin_2009} of ferrites and radius  $r=100~\mathrm{nm}$ yielding  $b=0.094~\mathrm{eV}$.} \ak{As discussed above in Sec.~\ref{sn:spin_photon}, we are not aware of any material for which all the material parameters required to quantify our proposed effect have been measured. Hence, we have taken typical experimental values available in the literature.} 

Also, we assume a dielectric function given by the Drude-Lorentz model with a single oscillator at angular frequency $\Omega$, oscillator strength $f_p$, loss factor $f_\gamma$, and background permittivity $\epsilon_{bg}$
\begin{equation}
	\epsilon = \epsilon_{bg} + \dfrac{f_p\Omega^2}{ \Omega^2-\omega ^2 -i f_\gamma\Omega  \omega}\;.
\end{equation}
By comparing the polarizability of a spherical particle in the quasistatic approximations with that of a system with background polarizability and a single dipole transition with dipole moment $d$, we obtain
\begin{align}
	d&=\dfrac{1}{\epsilon_{bg}+2}\sqrt{\dfrac{9  \epsilon_0 \hbar V f_p\Omega}{2 f_0}}\nonumber\\
	f_0&= \sqrt{1+\frac{f_p}{(\epsilon_{bg}+2)}-\dfrac{f_\gamma }{4}}\;.
\end{align}
For $\epsilon_{bg}=1$, $f_p=1$, $f_\gamma=0.01$, and $\Omega/(2\pi)=100~\mathrm{cm^{-1}}$\petros{, which correspond to typical values of iron garnets and cuprates}~\cite{grunberg_optical_1971,bernhard_-plane_2002,feneberg_ordinary_2018}, the dipole moment of each nanoparticle equals about $1.7~\mathrm{kD}$.

We note that our material parameters are within realistic range and in order to reach convergence, an $1600\times1600$-nanoparticle array is needed.

\section{Supplementary Note 6: Mean-field framework}\label{sn:mean_field}
According to the mean field approach, each spin fluctuates around its mean value \aknew{such that we may expand the spin operator as} $\mathbf{S}=\langle{\mathbf{S}}\rangle +\delta \mathbf{s}$, with \papnew{$ \delta s  \ll \langle S \rangle  $.} 
\aknew{With this approach,} for the $k$-th component we have $(S_{i;k})^2=(\langle{S_{i;k}}\rangle +\delta {s_{i;k})}^2$ \aknew{ $\approx \langle{S_{i;k}}\rangle^2 + 2 \langle{S_{i;k}}\rangle \delta {s_{i;k}} $ keeping up to first-order terms in $\delta {s_{i;k}}$. Thus, a} single term of Eq.~(4) in the main text takes the form
\begin{align}
	(S_{j;k})^2\tilde{\Lambda}_{j,j^\prime}^{kk^\prime}(S_{j^\prime;k^\prime})^2&\approx  -3 \langle{S_{j;k}}\rangle^2\tilde{\Lambda}_{j,j^\prime}^{kk^\prime}\langle{S_{j^\prime;k^\prime}}\rangle^2 + 2\langle{S_{j;k}}\rangle^2\tilde{\Lambda}_{j,j^\prime}^{kk^\prime}\langle{S_{j^\prime;k^\prime}}\rangle S_{j;k^\prime} \nonumber\\&+ 2S_{j;k}\langle{S_{j;k}}\rangle\tilde{\Lambda}_{j,j^\prime}^{kk^\prime}\langle{S_{j^\prime;k^\prime}}\rangle^2\;,
\end{align}
\aknew{where we have further employed $\delta {s_{i;k}} = S_{i;k} - \langle{S_{i;k}}\rangle$.} Assuming that the mean value of all spins is the same, Eq.~(4) of the main text can be cast in the form
\petros{\begin{align}\label{eq:H_meanfield}
		H_\mathrm{MF} &=\dfrac{3}{S^4}\langle\mathbf{S}\rangle^2\cdot\boldsymbol{\Lambda}\cdot\langle\mathbf{S}\rangle^2 -\dfrac{1}{S}\sum_{j}\mathbf{h}_{j}\cdot\mathbf{S}_j\;,
\end{align}}
where $\langle\mathbf{S}\rangle^2 \aknew{\equiv} (\langle S_x \rangle ^2,\langle S_y \rangle ^2,\langle S_y \rangle ^2)$, $\boldsymbol{\Lambda}=\sum_{j,j^\prime}\tilde{\boldsymbol{\Lambda}}_{j,j^\prime}$, \petros{$ \mathbf{h}_{j} = 2\boldsymbol{\sigma} \left(\boldsymbol{\Lambda}_{j}+\boldsymbol{\Lambda}_{j}^\mathrm{T}\right) \langle\mathbf{S}\rangle/S^3$},  $\boldsymbol{\Lambda}_{j}=\sum_{j^\prime}\tilde{\boldsymbol{\Lambda}}_{j,j^\prime}$,  and $\boldsymbol{\sigma}=\mathrm{diag}(\langle S_x \rangle ^2,\langle S_y \rangle ^2,\langle S_y \rangle ^2)$.
We note that $H_\mathrm{S}$ has been neglected, which can be realized by a Zeeman term for switched off external magnetic field.
It can be shown that for the hexagonal/square arrays in the $x$-$y$ plane, the mean-field coupling tensor is diagonal and then
\petros{\begin{align}
		\mathbf{h}_j = \dfrac{4}{S^3}\left(\Lambda_{j;x} \langle S_x \rangle ^3 \widehat{\mathbf{x}} + \Lambda_{j;y} \langle S_y \rangle ^3 \widehat{\mathbf{y}} + \Lambda_{j;z}\langle S_z \rangle ^3 \widehat{\mathbf{z}}\right)\;, \label{eq:mean_field_h}
	\end{align}
	$\Lambda_{k;j}=\sum_i \tilde{\Lambda}_{j,j^\prime}^{kk}$ for $k=x,y,z$. 
	
	Interestingly, the mean-field Hamiltonian, Eq.~\eqref{eq:H_meanfield}, has the same form as in the case of conventional ferromagnetism, in which $\mathbf{h}_j$ can be viewed as an, effective, internal molecular magnetic field. 
	Following the standard procedure of the mean-field theory\aknew{~\cite{coey_magnetism_2001,Gurevich1996}}, the self-consistent equation for the expectation value of the $k$-th component of the spin operator for the $j$-th nanoparticle is
	\begin{align}
		\langle S_{k} \rangle &=\dfrac{h_k}{h}SB_S{(\beta h)}\;,
	\end{align}
	where $B_S(x) = \dfrac{2S+1}{2S}\coth\left[\dfrac{(2S+1)x}{2S}\right]-\dfrac{1}{2S}\coth\left(\dfrac{x}{2S}\right) $ is the Brillouin function and $h=4 \sqrt{\sum_{k}(\Lambda_{k}\langle{S}_k\rangle^3)^2}$.  
	We note that the $j$ dependence has been dropped because all spins are identical.
	
	In the limit of $S\to\infty$ the Brillouin function reduces to the Langevin function, $L(x)=-1/x+\coth(x)$, and the self-consistent equation reads
	\begin{align}\label{eq:self_langevin}
		\dfrac{  \langle S_{x} \rangle}{S}=\dfrac{h_k}{h}\left[-\dfrac{1}{\beta h}+\coth\left( \beta h\right)\right]\;.
	\end{align}

	The Helmholtz free energy, $ F = U-TS$, reads
	\begin{equation}\label{eq:Fj}
		F=\dfrac{N}{\beta}\left\{\dfrac{3\beta}{S^4}{\sum_{k}(\Lambda_{k}^2\langle{S}_k\rangle^4)} - \ln \left[\cosh\left(\beta h\right) +\coth \left(\dfrac{\beta h}{2S}\right) \sinh\left(\beta h\right)\right]\right\}\;.
\end{equation}}

\section{Supplementary Note 7: Validating the long-wavelength approximation}\label{sec:wavelength}

\aknew{The effective spin-spin coupling [Eq.~(\ref{eq:coupling_ij_final})] is obtained by carrying out a} \papnew{contour integration,} \aknew{with the contour including the entire real frequency axis [see Eqs.~(\ref{eq:xi_gf}) - (\ref{eq:coupling_ij_final})]. 
	In practice, the long-wavelength or dipole approximation employed in our treatment of the photon-phonon coupling is valid up to frequencies with the corresponding optical wavelength much larger than the nanoparticle size. 
	This imposes a physical restriction on the quantity that is considered $\infty$ in the frequency integral Eq.~(\ref{eq:xi_gf}). 
	Here, we demonstrate that imposing the appropriate physical frequency cut-off, instead of carrying out the integral to $\infty$, still yields the same result for the physical systems considered herein. 
	Such considerations and results are common in various physical phenomena when a formally infinite value needs to be replaced by a high cut-off governed by certain physical constraints.
	
	We consider the free-space dyadic Green's function}
\begin{equation}\label{eq:freegf}
	\papnew{\tilde{\mathbf{G}}_0(\mathbf{r}_i,\mathbf{r}_j,\omega) = \frac{\mathrm{e}^{i k r}}{4\pi r}
		\left[\tilde{\mathbf{I}} - \widehat{\mathbf{r}} \otimes \widehat{\mathbf{r}}+
		\frac{(i k r - 1)}{k^2 r^2} (\tilde{\mathbf{I}} - 3 \widehat{\mathbf{r}} \otimes\widehat{\mathbf{r}}) \right]},
\end{equation}
\aknew{where $k = \omega/c$ is the optical wavenumber,} \papnew{ $\tilde{\mathbf{I}}$ is the $3\times3$ identity matrix$, \mathbf{r} = \mathbf{r}_i-\mathbf{r}_j$, and $\widehat{\mathbf{r}}=\mathbf{r}/r$.
	
	We are interested in the integral
	\begin{equation}\label{eq:I0_integral}
		\tilde{\mathbf{I}}_0=\dfrac{1}{\pi\epsilon_0c^2}\int_0^\infty \mathrm{d}\omega \omega \Im\tilde{\mathbf{G}}_0(\mathbf{r}_i,\mathbf{r}_{j},\omega)\;,
	\end{equation}
	which, using contour integration, can be calculated analytically (see Supplementary Note~\ref{sn:mqed})
	\begin{equation}\label{eq:I0_analytical}
		\tilde{\mathbf{I}}_0= \dfrac{1}{2\epsilon_0c^2} \left[\omega^2\tilde{\mathbf{G}}_0(\mathbf{r}_i,\mathbf{r}_{j},\omega)\right]_{\omega=0}\;.
\end{equation}}
\aknew{By inserting Eq.~(\ref{eq:freegf}) into Eq.~(\ref{eq:I0_analytical}) yields the kernel of the free-space electrostatic dipole-dipole interaction energy as per the expectations}\papnew{
	\begin{equation}\label{eq:free_space_dd_SM}
		\tilde{\mathbf{I}}_0 = \frac12 \frac{3  \widehat{\mathbf{r}} \otimes  \widehat{\mathbf{r}} - \aknew{\tilde{\mathbf{I}}}}{4\pi \epsilon_0 r^3},
\end{equation}}
\aknew{where the factor $1/2$ accounts for each pair of dipole appearing
	twice in the sum over $i$ and $j$.

	We next evaluate the integral in Eq.~(\ref{eq:I0_integral}) with an exponential cutoff at wavelength $\lambda_c$, i.e.,
	a cut-off frequency of $\omega_c = 2\pi c/\lambda_c$. 
	To simplify the expressions, we multiply the denominator of Eq.~(\ref{eq:free_space_dd_SM}) by $8\pi\epsilon_0 r^3$,
	yielding}
\papnew{
	\begin{align}
		8\pi\epsilon_0 r^3 \tilde{\mathbf{I}}_{\omega_c}
		= & ~ \dfrac{8 r^3}{c^2} \int_0^\infty \mathrm{d}\omega \omega \Im{\mathbf{G}_0}(\mathbf{r}_i,\mathbf{r}_{j},\omega) \mathrm{e}^{-\omega^2/\omega_c^2} \nonumber \\
		= & ~ 4 \pi^{5/2} r_\lambda^3 e^{-\pi^2 r_\lambda^2} \left(\tilde{\mathbf{I}} -  \widehat{\mathbf{r}}\otimes \widehat{\mathbf{r}}\right) + \nonumber \\ 
		& \left(1 - \operatorname{erfc}(\pi r_\lambda) - 2\sqrt{\pi} r_\lambda e^{-\pi^2 r_\lambda^2}\right) \left(3\widehat{\mathbf{r}}\otimes\widehat{\mathbf{r}} - \tilde{\mathbf{I}}\right),
	\end{align}
	where we have introduced $r_\lambda = r/\lambda_c$ for simplicity. 
	In the limit $\lambda_c\to 0$ (i.e., $\omega_c \to \infty$), this clearly recovers Eq.~(\ref{eq:free_space_dd_SM}). }
\aknew{Interestingly, this convergence with the increasing cut-off frequency $\omega_c$ is very fast, as all terms apart from the one obtained in Eq.~(\ref{eq:free_space_dd_SM}) are suppressed by $\mathrm{e}^{-\pi^2 r_{\lambda}^2}$ (which is also the asymptotic behavior of $\operatorname{erfc}$). 
	For $r_\lambda = 1$, the relative error is on the level of $0.4\%$, and for $r_\lambda = 2$, it is already below $10^{-14}$. 
	This implies that for a given distance $r$ between points, a cutoff wavelength similar to that distance is sufficient to obtain a fully converged result. 
	Since the spatial extent of our considered nanoparticles can at most be of the order of their separation (and is typically significantly smaller), this implies that the dipole approximation works well. 
	The formal extension of the frequency integral to an upper limit of infinity without going beyond the dipole approximation is thus well-justified. 
	We note that while we have here explicitly demonstrated the case of free space, the same arguments are expected to apply in general environments.
	
}

\end{document}